\documentclass[english,notitlepage,nofootinbib]{revtex4-1}
\pdfoutput=1
\usepackage[T1]{fontenc}
\usepackage[latin9]{inputenc}
\usepackage{color}
\usepackage{verbatim}
\usepackage{amsmath}
\usepackage{stackrel}
\usepackage{graphicx}
\usepackage{esint}
\usepackage{babel}
\usepackage{url}
\usepackage{hyperref}

\def\gs{\mathrel{
   \rlap{\raise 0.511ex \hbox{$>$}}{\lower 0.511ex \hbox{$\sim$}}}}
\def\ls{\mathrel{
   \rlap{\raise 0.511ex \hbox{$<$}}{\lower 0.511ex \hbox{$\sim$}}}}

\begin{document}

\title{Coherent Neutrino-Nucleus Scattering and New Neutrino Interactions}

\date{\today}

\author{Manfred Lindner, Werner Rodejohann, Xun-Jie Xu }

\affiliation{Max-Planck-Institut f\"ur Kernphysik, Postfach 103980, D-69029 Heidelberg,
Germany}
\begin{abstract}
\noindent
We investigate the potential to probe new neutrino physics with
future experiments measuring coherent neutrino-nucleus scattering.
Experiments with high statistics should become feasible soon and
allow to constrain parameters with unprecedented precision. Using a
benchmark setup for a future experiment probing reactor neutrinos,
we study the sensitivity on
neutrino non-standard interactions and new exotic neutral currents (scalar, tensor, etc).
Compared to Fermi interaction, percent and permille level strengths
of the new interactions can be probed,
superseding for some observables
the limits from future neutrino oscillation experiments by up to two
orders of magnitude.

\end{abstract}
\maketitle

\section{Introduction}

Coherent neutrino-nucleus scattering (C$\nu N$S)
\cite{Freedman:1973yd,Freedman:1977xn,Drukier:1983gj} is a tree
level process that is
predicted by the Standard Model, but has not yet been observed. While being
conceptually highly interesting and allowing measurements of electroweak
observables at low momentum transfer, the process is also of
phenomenological importance for future dark matter direct detection experiments
\cite{Billard:2013qya}. Moreover, it holds the potential
to probe new neutrino physics \cite{Barranco:2005yy,Dutta:2015vwa,Dutta:2015nlo,Papoulias:2015vxa}, which is the main focus of this paper.

In C$\nu N$S, low energy neutrinos interact with the
protons and neutrons in the nuclei coherently, which significantly enhances the
cross section. While large fluxes of neutrinos are available from nuclear
research or commercial reactors, the recoil energy of the nuclei
is difficult to detect since it is very low.
However, prompted partly by developments in dark matter direct detection experiments,
modern low-threshold detectors make the detection of C$\nu N$S technically
feasible \cite{Scholberg:2005qs,Wong:2005vg}. Combined with smart shielding techniques,
high-rate and low-background experiments are possible\footnote{See, for instance, Ref.\  \cite{Akimov:2015nza,Wong:2016lmb,Kerman:2016jqp,Soma:2014zgm,Anderson:2011bi} for recent studies.}.
Future C$\nu N$S experiments may thus provide precision
test of neutrino interactions in the Standard Model and
strong constraints on new physics related to neutrinos.


In this paper, we will study the sensitivities of C$\nu N$S on possible
new neutrino interactions, mainly assuming Germanium detectors
with sub-keV threshold, detecting reactor antineutrinos. For illustration, we will assume
 values of the experimental parameters within reach
of current technology\footnote{See e.g.\ https://indico.mpp.mpg.de/event/3121/session/3/contribution/18/material/slides/0.pdf for details.}.
To make our study applicable to various new physics models, we will adopt a
model-independent approach, only considering the low energy effective operators of neutrinos
and quarks. This includes not only the widely-discussed conventional
Non-Standard Interactions (NSI) \cite{Davidson:2003ha} which are
in (chiral) vector form, but also more exotic interactions that could
be in scalar or tensor form.
 What distinguishes this paper from previous
studies of the potential implications of coherent scattering
\cite{Barranco:2005yy,Dutta:2015vwa,Dutta:2015nlo,Papoulias:2015vxa,Cerdeno:2016sfi}, is the inclusion of
such exotic interactions, and a comparative study on how different experimental details
(such as energy threshold or neutrino flux uncertainty) influence the sensitivity
on new physics.

The paper is organized as follows. We start by introducing C$\nu N$S in
the Standard Model in Sec.\ \ref{sec:scat-in-SM}. Then we study the effect
of new physics on  C$\nu N$S, based on effective operators of neutrinos
and quarks, which can be divided into two cases, the conventional
NSI in Sec.\ \ref{sec:NSI} and exotic neutral currents in Sec.\ \ref{sec:Exotic}.
In Sec.\ \ref{sec:fit}, we consider a benchmark setup for a C$\nu N$S experiment
and perform $\chi^{2}$-fit
on parameters from the Standard Model, NSI and exotic neutral currents to study
the sensitivities of such an experiment on them. We conclude
in Sec.\ \ref{sec:Conclusion}. Details on the calculation of the cross section
with both spin-0 and spin-$1/2$ nuclei are delegated
to Appendix \ref{sec:Cross-section} and \ref{sec:What-if}.
Some useful relations connecting the fundamental coupling constants of exotic
neutral currents to the effective parameters in C$\nu N$S are given in
Appendix \ref{sec:formfactor}.

\section{\label{sec:scat-in-SM}Coherent neutrino-nucleus scattering in the
Standard Model}

\subsection{Cross Section}

In the Standard Model (SM), the Neutral Current (NC) interaction enables
low energy neutrinos with $E_\nu\ls 50$ MeV (corresponding to length scales
of $\gs 10^{-14}$ m) to interact  coherently with protons and neutrons in a
nucleus, which significantly enhances the cross section for a
large nucleus. For a nucleus at rest with $Z$ protons and $N$
neutrons, the coherent cross section \cite{Freedman:1973yd,Freedman:1977xn,Papoulias:2015vxa}
(see Appendix \ref{sec:Cross-section}) is given by
\begin{equation}
\frac{d\sigma}{dT}=\frac{\sigma_{0}^{{\rm SM}}}{M}\left(1-\frac{T}{T_{\max}}\right),\label{eq:coh-34}
\end{equation}
where $\sigma_{0}^{{\rm SM}}$ is defined as
\begin{equation}
\sigma_{0}^{{\rm SM}}\equiv\frac{G_{F}^{2}\left[N-(1-4s_{W}^{2})Z\right]^{2}F^{2}(q^{2})M^{2}}{4\pi}\,.\label{eq:coh-33}
\end{equation}
Here $G_{F}$, $s_{W}=\sin\theta_{W}$, and $M$ are the Fermi constant,
the Weinberg angle, and the mass of the nucleus, respectively. Since at low energies
$s_{W}^{2}\approx0.238$ \cite{Erler:2004in}, we have $N-(1-4s_{W}^{2})Z$
$\approx N-0.045Z$, which implies that the cross section is dominated
by the neutron number; $F(q^{2})$ is the form factor of the nucleus
and its coherent limit ($q^{2}\rightarrow0$) is $1$. For higher
energies, due to  loss of coherence, it will be smaller than
$1$ (for a recent quantitative study, see  Ref.\ \cite{Kerman:2016jqp}).
The recoil energy $T$ of the nucleus has a maximal value $T_{{\rm max}}$,
determined by the initial neutrino energy $E_{\nu}$ and the nucleus mass $M$:
\begin{equation}
T_{{\rm max}}(E_{\nu})=\frac{2E_{\nu}^{2}}{M+2E_{\nu}}\,.\label{eq:coh-29-1}
\end{equation}
For new physics beyond the SM, both Eq.\,(\ref{eq:coh-34}) and Eq.\,(\ref{eq:coh-33})
could be modified but Eq.\,(\ref{eq:coh-29-1}) still holds since
it is determined purely from relativistic kinematics.

Eq.\ (\ref{eq:coh-33})
was derived under the assumption that the nucleus is a spin-0 particle
\cite{Freedman:1973yd} (see also Appendix \ref{sec:Cross-section}
of this paper). However, this is not always true because a nucleus
with odd $A=N+Z$ is a fermion, examples are $^{73}{\rm Ge}$ or
$^{131}{\rm Xe}$.
In Appendix \ref{sec:What-if}, we calculate the simplest non-zero
case, spin-$1/2$. It turns out that the difference is small,
given by
\begin{equation}
\left.\frac{d\sigma}{dT}\right|_{{\rm {\rm spin}}=\frac{1}{2}}=\left.\frac{d\sigma}{dT}\right|_{{\rm {\rm spin}}=0}+\frac{\sigma_{0}^{{\rm SM}}}{M}\frac{T^{2}}{2E_{\nu}^{2}}\,.\label{eq:coh-41}
\end{equation}
Thus, the only difference is a term proportional to $T^{2}/E_{\nu}^{2}$,
which is usually negligible in the coherence scattering process. In
principle the nucleus could also be some higher spin particle but
based on Eq.\ (\ref{eq:coh-41}) it is reasonable to deduce that the
difference should be suppressed for a large nucleus.

\subsection{\label{sub:Detection}Detection}

Note that the recoil energy $T$ is the only measurable effect of
coherent neutrino scattering. Depending on the type of detectors,
 the method to measure $T$ is very different. We will focus
here on Germanium detectors which  measure
 the ionization energy $I$, which is a fraction of
the deposited recoil energy $T$. The fraction is defined as the quenching factor
$Q=I/T$, typically within 0.15 to 0.3 for sub-keV recoil energies
(see e.g.\ Figs.\ 5 and 7 in Ref.\ \cite{Barker:2012ek}).
The quenching factor at sub-keV energies is not well known due to
 lack of experimental data. In typical models like the one
proposed by Lindhard et al.\ \cite{Lindhard}, the recoil energy
depends on $Q$, so $I=TQ(T)$ would be a (not necessarily
linear) function of $T$.
However, no matter what the exact form of the function
$I(T)$ would be, once $I$ is measured, it can be converted to $T$,
provided that this function has been theoretically calculated \cite{Barker:2012ek}
or experimentally measured\footnote{One approach to measure the quenching factor is
to use neutron scattering, as performed in the CDEX-TEXONO collaboration above keV energies.
For more details see https://wwwgerda.mpp.mpg.de/symp/20\_Ruan.pdf.
In the future sub-keV measurements will be performed. }. We assume that the
quenching factor can be measured precisely in the future, and thus
use the recoil energy $T$ rather than the  ionization energy $I$.
All the results in our
paper can be simply converted from the $T$-dependence to the
$I$-dependence, provided that the function $I(T)$ is determined.

\begin{figure}
\centering

\includegraphics[width=8cm]{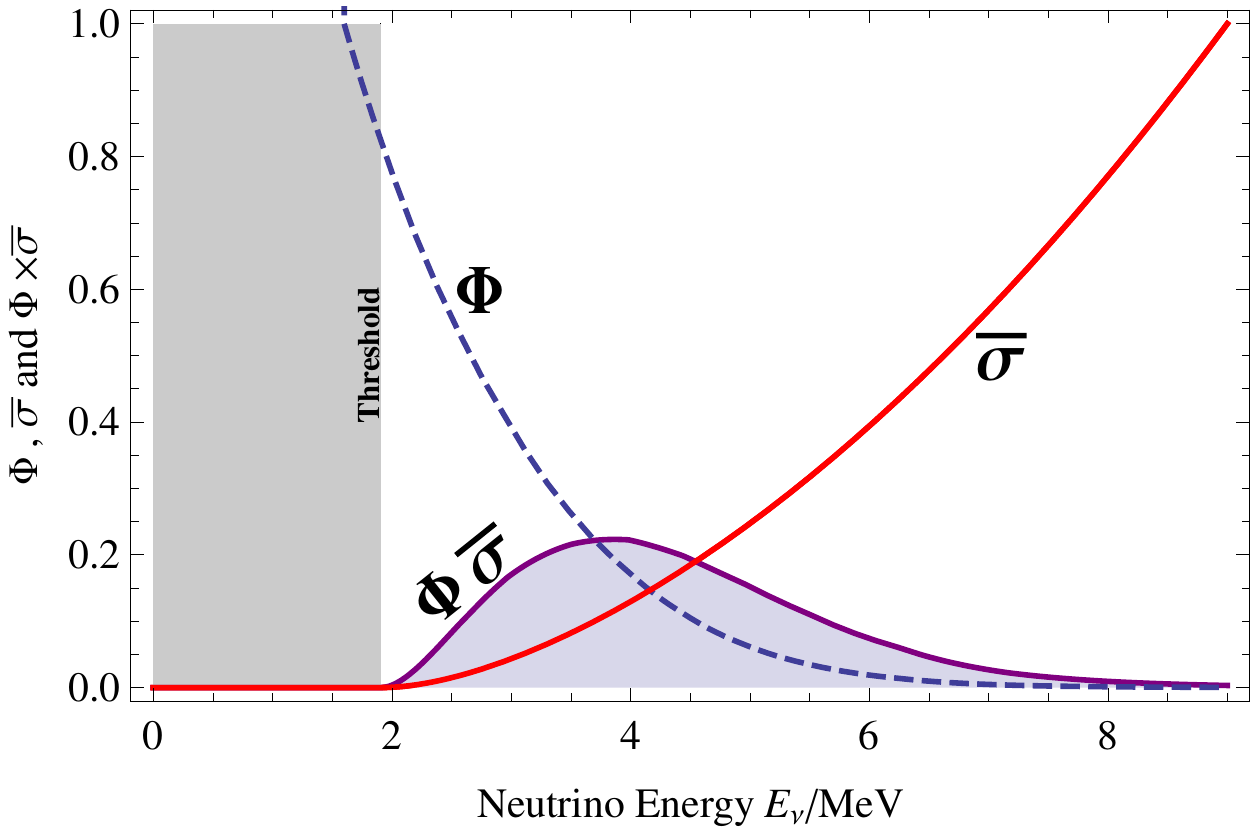}

\caption{\label{fig:flux-sigma}A typical reactor neutrino flux $\Phi$, reduced $\nu-N$
scattering cross section $\bar{\sigma}$ and their product. Units are arbitrary.}
\end{figure}

Generally for all types of detectors there is a detection threshold
on $T$, denoted as $T_{{\rm th}}$.  Therefore, for a given $E_{\nu}$
the recoil energy $T$ of detected events should be within the range
$T_{{\rm th}}\leq T\leq T_{{\rm max}}$ and the measurable reduced
total cross section is
\begin{equation}
\bar{\sigma}(T_{{\rm th}},\thinspace E_{\nu})\equiv\stackrel[T_{{\rm th}}]{T_{{\rm max}}}{\int}\frac{d\sigma}{dT}dT=\sigma_{0}^{\rm SM}\frac{\left(T_{\max}-T_{\text{th}}\right){}^{2}}{2MT_{{\rm max}}} \,.\label{eq:coh-31}
\end{equation}
Due to the threshold $T_{{\rm th}}$, low energy neutrinos are impossible
to detect if their energies are lower than
\begin{equation}
E_{\nu,{\rm th}}=\frac{1}{2}\left(\sqrt{2MT_{{\rm th}}+T_{{\rm th}}^{2}}+T_{{\rm th}}\right)
\approx \sqrt{\frac{M}{2}T_{{\rm th}}}\,.\label{eq:coh-37}
\end{equation}
For example, if $T_{{\rm th}}=0.1$ keV \cite{Wong:2016lmb} then
neutrinos should have $E_{\nu}>E_{\nu,{\rm th}}\approx2$ MeV in order
to be detected in a Ge detector.  On the other hand, if we consider
reactor neutrinos, the flux decreases exponentially at high energy.
Therefore there is a limited range of $E_{\nu}$ for detection.
To illustrate this, we plot in Fig.\ \ref{fig:flux-sigma}  the reduced cross section
$\bar{\sigma}$ {[}given by Eq.\ (\ref{eq:coh-31}){]}, a typical reactor
neutrino flux $\Phi$ and their product $\Phi\bar{\sigma}$,
which is essentially proportional to the event rate.
As Fig.\ \ref{fig:flux-sigma} shows, the product $\Phi\bar{\sigma}$ is
small at both low (2 MeV) and high (8 MeV) energies.

\begin{figure}
\centering

\includegraphics[width=8cm]{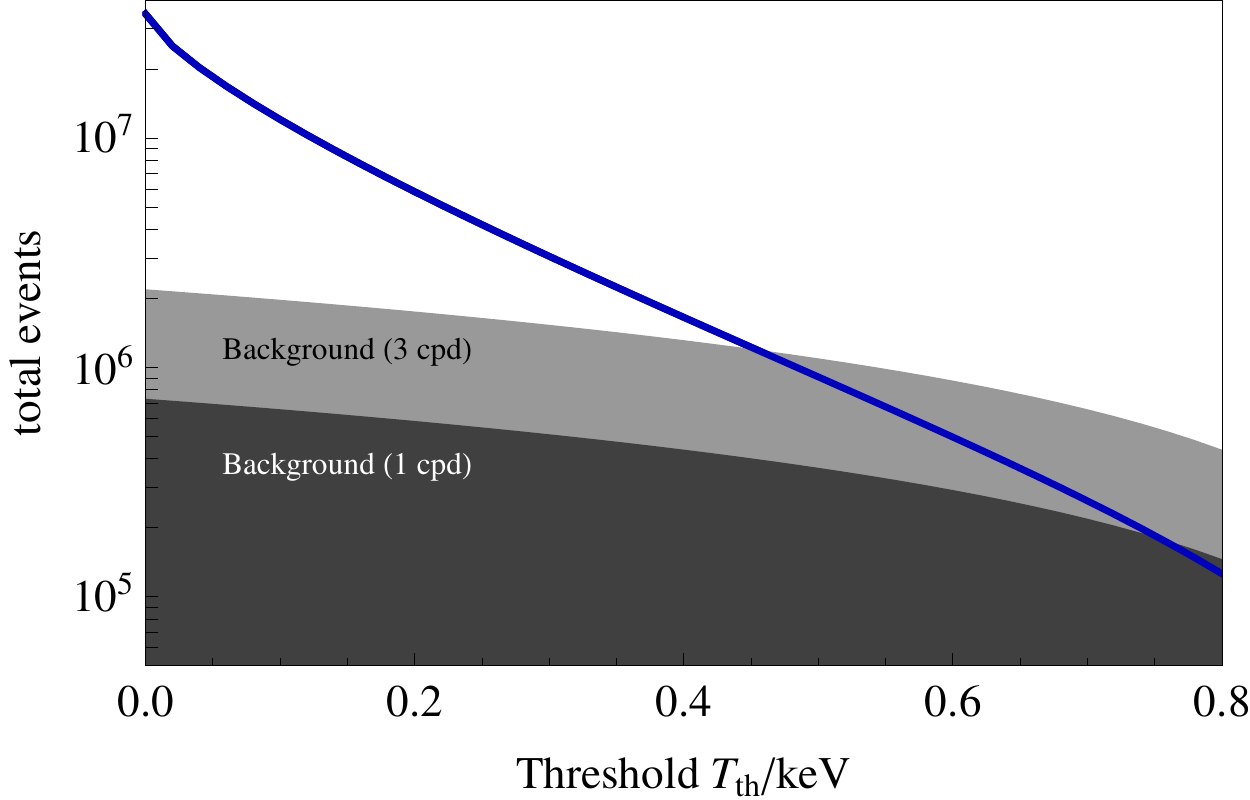}

\caption{\label{fig:threshold} Total number of events compared with background
(1 cpd = $1$ ${\rm day}^{-1}\thinspace{\rm kg}^{-1}\thinspace{\rm keV}^{-1}$).
The total number decreases significantly when the detection threshold
increases.  We assume a 100 kg Ge detector located 10 m
away from a reactor with 1 GW thermal power, taking data for five years. For zero threshold, the total number of events is $3.8\times10^{7}$.}
\end{figure}

From the above discussion it is clear that the total event number decreases
drastically when the detection threshold $T_{{\rm th}}$ is increased.
To show this, we compute the total event numbers with different
detection thresholds, plotted in Fig.\ \ref{fig:threshold}, where
one can see that the event number drops by 2 orders of magnitude
if $T_{{\rm th}}$ rises from $0.1$ keV to 0.8 keV. Therefore lowering the detection
threshold is very crucial in order to obtain large event numbers.
For this plot we have assumed
a 100 kg Ge detector located 10 m away from a 1 GW (thermal power)
reactor and taking data for five years.
For the neutrino flux $\Phi(E_{\nu})$, we have taken
the spectrum from a recent
theoretical calculation in Ref.\ \cite{Kopeikin:2012zz},
normalized to
$1.7\times10^{13}\thinspace\text{cm}^{-2}\thinspace{\rm s}^{-1}$
(corresponding to 10 m distance from the reactor). Those
values will serve as benchmark for our assumed future experiment, and can be used as a
definition of our assumed ``exposure'' of
\begin{equation}\label{eq:expo}
{\rm exposure } = 5 \, {\rm kg \cdot yr \cdot GW  \cdot m^{-2}}\,.
\end{equation}
In Fig.\ \ref{fig:threshold} we also show the effect of
an assumed constant background of 1 cpd and 3 cpd (1 cpd =
$1$ ${\rm day}^{-1}\thinspace{\rm kg}^{-1}\thinspace{\rm keV}^{-1}$).
The background may come from various sources, such as the intrinsic
radioactivity of the material in the detector, ambient radioactivity
near the nuclear reactor or cosmic rays. Estimation of the background
is very much involved and depends significantly on the details
of the detector. 
The GEMMA experiment \cite{Beda:2013mta} states a
background level of about 2 cpd and the TEXONO collaboration is aiming
at developing a Ge detector with a background of 1 cpd \cite{Wong:2005vg}.
Note however that the mentioned background numbers apply to somewhat different
energy scales and different background sources. Taking into account the low
background levels that various double beta decay and dark matter direct detection exeriments
have reached, plus noting the developements on active shielding at
shallow depth \cite{Heusser:2015ifa}, we estimate that such low background
rates can be achieved.

In reality, not only the total event number but also the
distribution of  events will be measured, giving us a spectrum
with respect to the recoil energy $T$. The spectrum provides more
information than the total event number. The advantage to exploit
the spectrum is that it is not influenced strongly by many uncertainties
such as  the flux normalization, the distance and fiducial mass of
the detector, the form factor, etc. All those effects can be described roughly
by an overall factor that enhances/reduces the total event number.

\begin{figure}
\centering

\includegraphics[width=8cm]{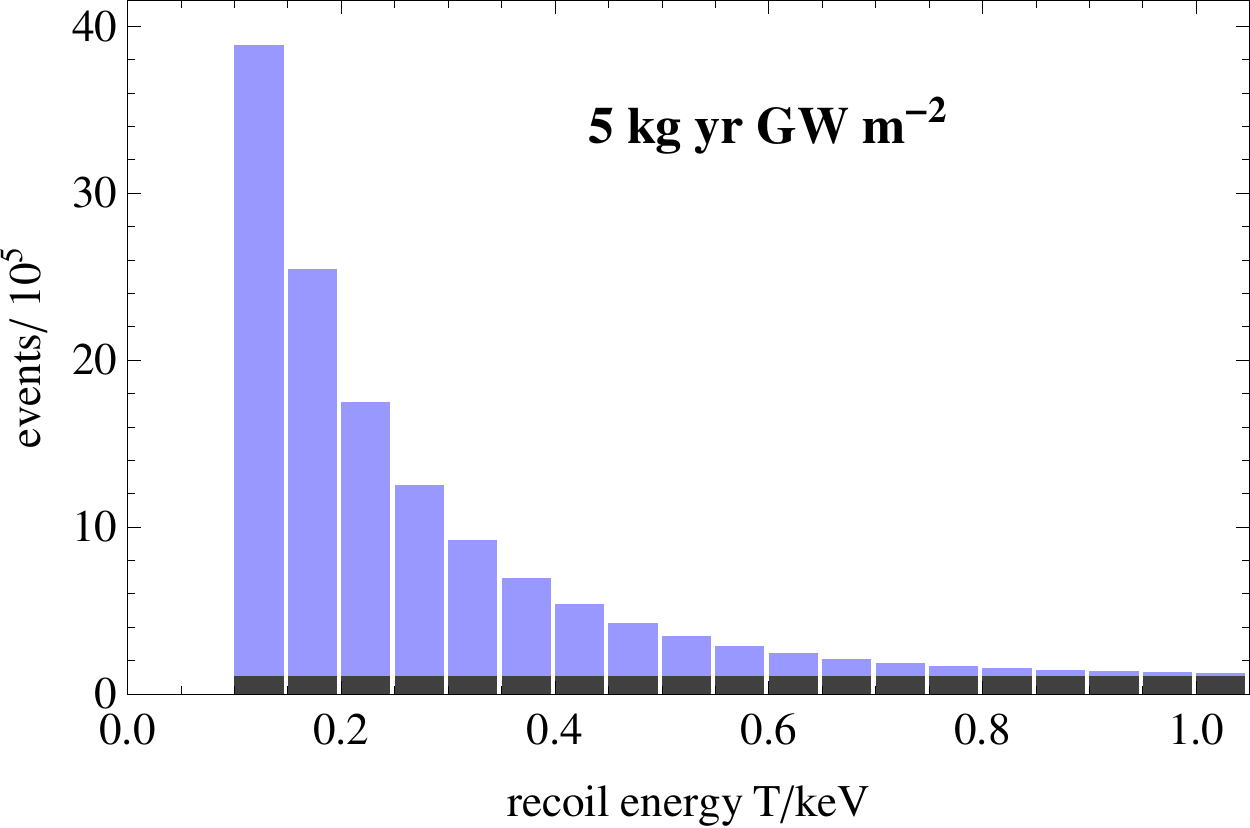}

\caption{\label{fig:ge1kg1yr}Expectation of event numbers in a 100 kg ${\rm Ge}$
detector running for 5 years, assuming a total flux of reactor neutrinos
of $1.7\times10^{13}\thinspace\text{cm}^{-2}\thinspace{\rm s}^{-1}$.
The background (black) is assumed to be 3 cpd.}
\end{figure}

If the events are  conservatively  counted in many $T$-bins, the $i$-th bin with width $\Delta T$ starting
from $T_{i}$, then the expectation of
the event number $N_{i}$ in the $i$-th bin is
\begin{equation}
N_{i}=\Delta t\,N_{{\rm Ge}}\frac{\sigma_{0}^{{\rm SM}}}{M}\int_{T_{i}}^{T_{i}+\Delta T}dT\int_{0}^{8\thinspace{\rm MeV}}dE_{\nu}\Phi(E_{\nu})f^{\rm SM}(T,E_{\nu})\,.\label{eq:coh-71}
\end{equation}
  Here $N_{{\rm Ge}}$ is the number of Ge nuclei\footnote{Natural Germanium consists of $^{70}{\rm Ge}$ ($20.52\%$), $^{72}{\rm Ge}$
(27.45\%), $^{73}{\rm Ge}$ (7.76\%), $^{74}{\rm Ge}$ (36.52\%) and
$^{76}{\rm Ge}$ (7.75\%). Here we take $A=72.6$ in average. { 
Note that spin-dependent axial couplings in the Standard Model lead to smaller coherence
factors depending on the spin of the nucleus, not on $N$ or $Z$ as the vector interaction that
gives the leading contribution, see Appendix \ref{sec:What-if}. This will be a permille effect,
see \cite{Cerdeno:2016sfi}}.} in the detector and $\Delta t$ is the running time of detection,
taken as 5 years. The neutrino flux $\Phi(E_{\nu})$ has been taken from
\cite{Kopeikin:2012zz}, and the dimensionless
function $f^{{\rm SM}}(T,E_{\nu})$ is defined
as {[}see Eq.\ (\ref{eq:coh-34}){]}
\begin{equation}
f^{{\rm SM}}(T,E_{\nu})=\begin{cases}
1-\frac{T}{T_{\max}(E_{\nu})} & \thinspace{\rm for\thinspace}T\leq T_{\max}\\
0 & \thinspace{\rm for\thinspace}T>T_{\max}
\end{cases}.\label{eq:coh-72}
\end{equation}
Note that when new physics beyond the SM is involved, one only needs
to modify $\sigma_{0}^{{\rm SM}}$ in Eq.\ (\ref{eq:coh-71}) and
$1-\frac{T}{T_{\max}(E_{\nu})}$ in Eq.\ (\ref{eq:coh-72}) according
to the new physics.
Taking the flux from Ref.\ \cite{Kopeikin:2012zz} and setting the
background at constant 3 cpd, the event numbers computed according to Eq.\
(\ref{eq:coh-71}) are  presented in Fig.\ \ref{fig:ge1kg1yr} as a function of $T$.

We should mention here that the calculation of the reactor neutrino
flux is very complicated. Though a lot of effort was spent to
calculate the flux in the literature
(see e.g.\ \cite{Kopeikin:2012zz,Achkar:1996rd,Schreckenbach:1985ep,Mention:2011rk,Huber:2011wv}
and references therein), so far a very precise result is
lacking,  especially for neutrino energies below 2 MeV where the
error could be large as 7\%. The best understood range is from 2 MeV
to 6 MeV, but still with 3\% error. Recently,
measurements from the RENO \cite{Seo:2014xei,RENO:2015ksa}, Daya
Bay \cite{An:2015nua} and Double Chooz \cite{Abe:2015rcp} experiments
showed disagreement
with the theoretical calculation around 5 MeV, the infamous 5 MeV
bump. Its observation implies that we might have not fully understood
the reactor neutrino flux. A particle physics origin of the bump
seems very unlikely.
In the next few years, both the theoretical
understanding and experimental measurement will be significantly improved
\cite{Buck:2015clx,Giunti:2016elf,Huber:2016xis} so that the flux will be known more precisely
and also the issue of the 5 MeV bump
will be resolved once our assumed future C$\nu N$S experiment is running. Anyway,
 the sensitivities of coherent
$\nu-N$ scattering on new physics depend very little on the
presence of the bump.
A quantitative study on the influence of the 5 MeV bump is presented below.

\section{\label{sec:NSI}Non-Standard Interactions
in Coherent $\nu-N$ Scattering}

Coherent $\nu-N$ scattering could provide very strong constraints
on  neutrino Non-Standard Interactions (NSI).  Those have been widely studied
in the literature but so far the experimental constraints on some
of its parameters are still very poor (see the reviews \cite{Davidson:2003ha}
and \cite{Ohlsson:2012kf}), especially the couplings of neutrinos
to quarks.

In this work, only the neutrino-quark sector of NSI is relevant. The
Lagrangian is
\begin{equation}
{\cal L}\supset\frac{G_{F}}{\sqrt{2}}\sum_{q=u,d}\overline{\nu}_{\alpha}\gamma^{\mu}(1-\gamma^{5})\nu_{\beta}\left[\varepsilon_{\alpha\beta}^{qV}\overline{q}\gamma^{\mu}q+\varepsilon_{\alpha\beta}^{qA}\overline{q}\gamma^{\mu}\gamma^{5}q\right],\label{eq:coh-36}
\end{equation}
where $\alpha,\thinspace\beta$ are the three flavors of neutrinos, and
$\varepsilon_{\alpha\beta}^{qV}$, $\varepsilon_{\alpha\beta}^{qA}$
are the non-standard vector and axial-vector coupling constants,
respectively.
Interpreting the NSI terms in analogy to Fermi theory implies that the various
$\varepsilon$ are given by
\begin{equation}\label{eq:NSITeV}
\varepsilon \approx \frac{g_X^2}{g^2} \frac{M_W^2}{M_X^2}\,,
\end{equation}
i.e.\ are related to new interactions mediated (for
$\varepsilon \sim 0.1$) by TeV-scale particles with mass
$M_X$ ($g_X$ denotes a new coupling constant).
In neutrino oscillation experiments
long-range forces have a similar effect as matter-induced NSIs
\cite{Heeck:2010pg}. We note that such light mediators could strongly
affect the shape of the spectrum under study here, and thus distinguish both possibilities. 

When the NSI Lagrangian (\ref{eq:coh-36}) is added to the SM, the C$\nu N$S differential cross section is changed only by an overall
factor. For the SM, the differential cross section is given in Eq.\ (\ref{eq:coh-34})
which is proportional to $\sigma_{0}^{{\rm SM}}$ given by Eq.\ (\ref{eq:coh-33}).
For the NSI, following the calculation in Appendix \ref{sec:Cross-section},
it is straightforward to obtain the result, which is simply replacing
$\sigma_{0}^{{\rm SM}}$ with $\sigma_{0}^{{\rm NSI}}$, given by
\begin{equation}
\sigma_{0}^{{\rm NSI}}=\frac{G_{F}^{2}Q_{{\rm NSI}}^{2}F^{2}(q^{2})M^{2}}{4\pi}\,.\label{eq:coh-62}
\end{equation}
Here the modified weak charge $Q_{{\rm NSI}}$ is defined as
\begin{eqnarray}
Q_{{\rm NSI}}^{2} & \equiv & 4\left[N\left(-\frac{1}{2}+\varepsilon_{ee}^{uV}+2\varepsilon_{ee}^{dV}\right)+Z\left(\frac{1}{2}-2s_{W}^{2}+2\varepsilon_{ee}^{uV}+\varepsilon_{ee}^{dV}\right)\right]^{2}\nonumber \\
 &  & +4\sum_{\alpha=\mu,\tau}\left[N(\varepsilon_{\alpha e}^{uV}+2\varepsilon_{\alpha e}^{dV})+Z(2\varepsilon_{\alpha e}^{uV}+\varepsilon_{\alpha e}^{dV})\right]^{2}.\label{eq:coh-63}
\end{eqnarray}
Setting the $\varepsilon$ to zero gives back the result from Eq.\ (\ref{eq:coh-33}).
The axial vector couplings $\varepsilon_{\alpha\beta}^{qA}$ in Eq.\ (\ref{eq:coh-36})
do not appear in Eq.\ (\ref{eq:coh-63}) because of parity symmetry
being present in
large nuclei (see the discussion in Appendix \ref{sec:Cross-section}).
The cross section only depends on the vector couplings
$\varepsilon_{\alpha\beta}^{qV}$,
which for simplicity will be denoted by $\varepsilon_{\alpha\beta}^{q}$
henceforth. Even though this removes a lot of parameters, we are still
confronted with a six-dimensional parameter space,
\begin{equation}
\overrightarrow{\varepsilon}\equiv(\varepsilon_{ee}^{u},\thinspace\varepsilon_{ee}^{d},\thinspace\varepsilon_{\mu e}^{u},\thinspace\varepsilon_{\mu e}^{d},\thinspace\varepsilon_{\tau e}^{u},\thinspace\varepsilon_{\tau e}^{d})\,.\label{eq:coh-64}
\end{equation}
So far the best constraints \cite{Davidson:2003ha}
on $\varepsilon_{\alpha e}^{q}$ ($\alpha=e,\thinspace\tau$) come from
CHARM $\nu_{e}(\overline{\nu}_{e})N$
inelastic scattering \cite{Dorenbosch:1986tb}. The 3$\sigma$-limits
are
\begin{eqnarray}
-1.2 & < & \varepsilon_{ee}^{u}<0.8,\thinspace\label{eq:coh-65}\\
-0.7 & < & \varepsilon_{ee}^{d}<1.4,\thinspace\label{eq:coh-67}\\
-1.0 & < & \varepsilon_{\tau e}^{u}<1.0,\thinspace\label{eq:coh-68}\\
-1.0 & < & \varepsilon_{\tau e}^{d}<1.0,\thinspace\label{eq:coh-69}
\end{eqnarray}
assuming that for each bound only the corresponding coupling is non-zero.
As one can see, these bounds are typically of order one. For the $\mu$
flavor, the best constraints are from
$\mu^{-}\,{\rm Ti}\rightarrow e^{-}\,{\rm Ti}$
\cite{deGouvea:2000cf,Davidson:2003ha},
\begin{equation}
|\varepsilon_{e\mu}^{u}|,\thinspace|\varepsilon_{e\mu}^{d}|<1.4\times10^{-3},\thinspace(3\sigma).\label{eq:coh-66}
\end{equation}
This bound comes from
a 1-loop diagram including the four fermion vertex of
$|\varepsilon_{e\mu}^{q}|$.
As a consequence, the result depends on the scale $\Lambda$ of the underlying
UV  complete model (recall that NSI in Eq.\ (\ref{eq:coh-36})
are non-renormalizable).
In Ref.\ \cite{Davidson:2003ha} it is assumed
$\ln(\Lambda/m_{W})\approx1$ to obtain this bound.

\begin{figure}
\begin{minipage}{13cm}

\includegraphics[width=6cm]{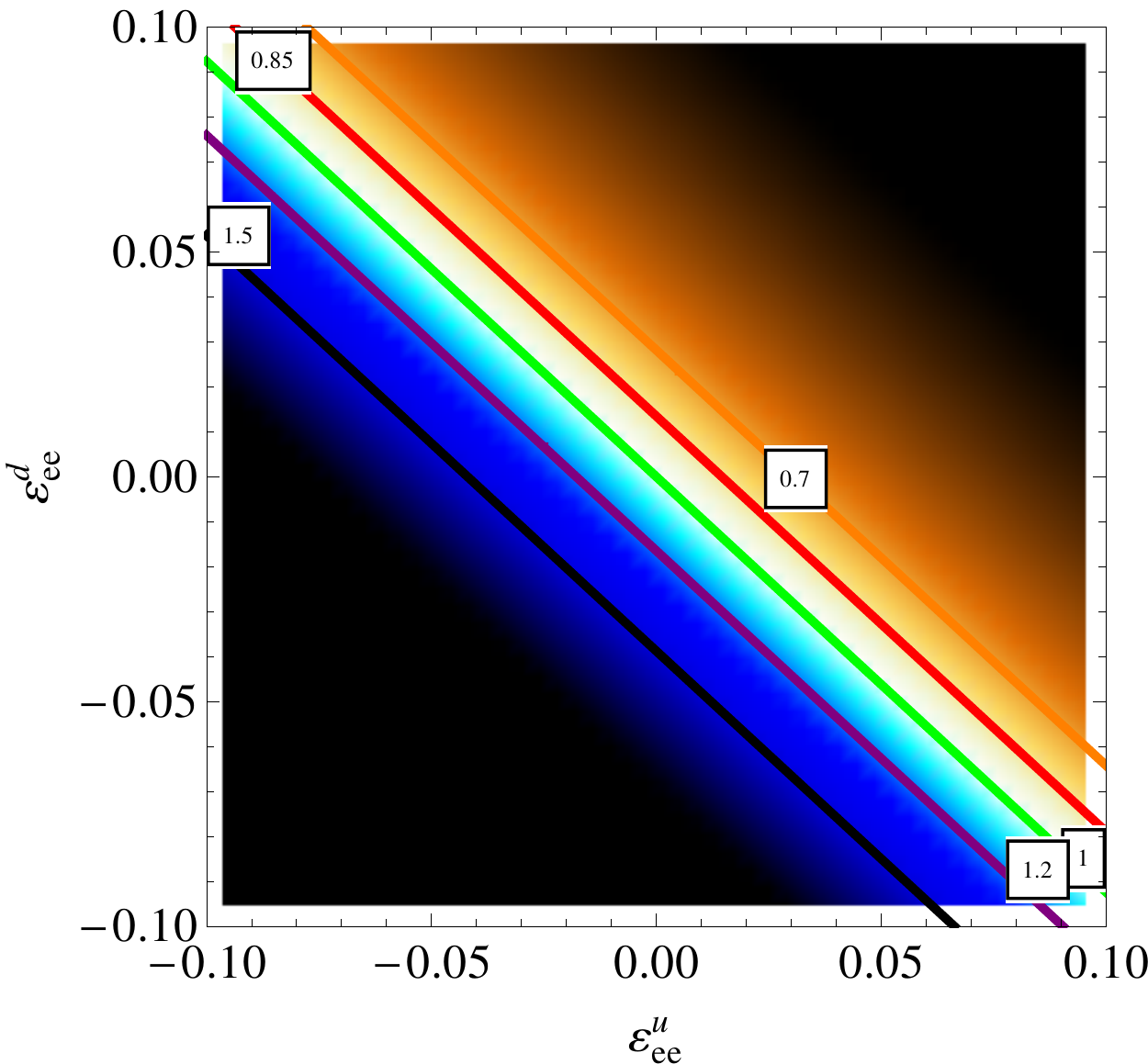}\includegraphics[width=6cm]{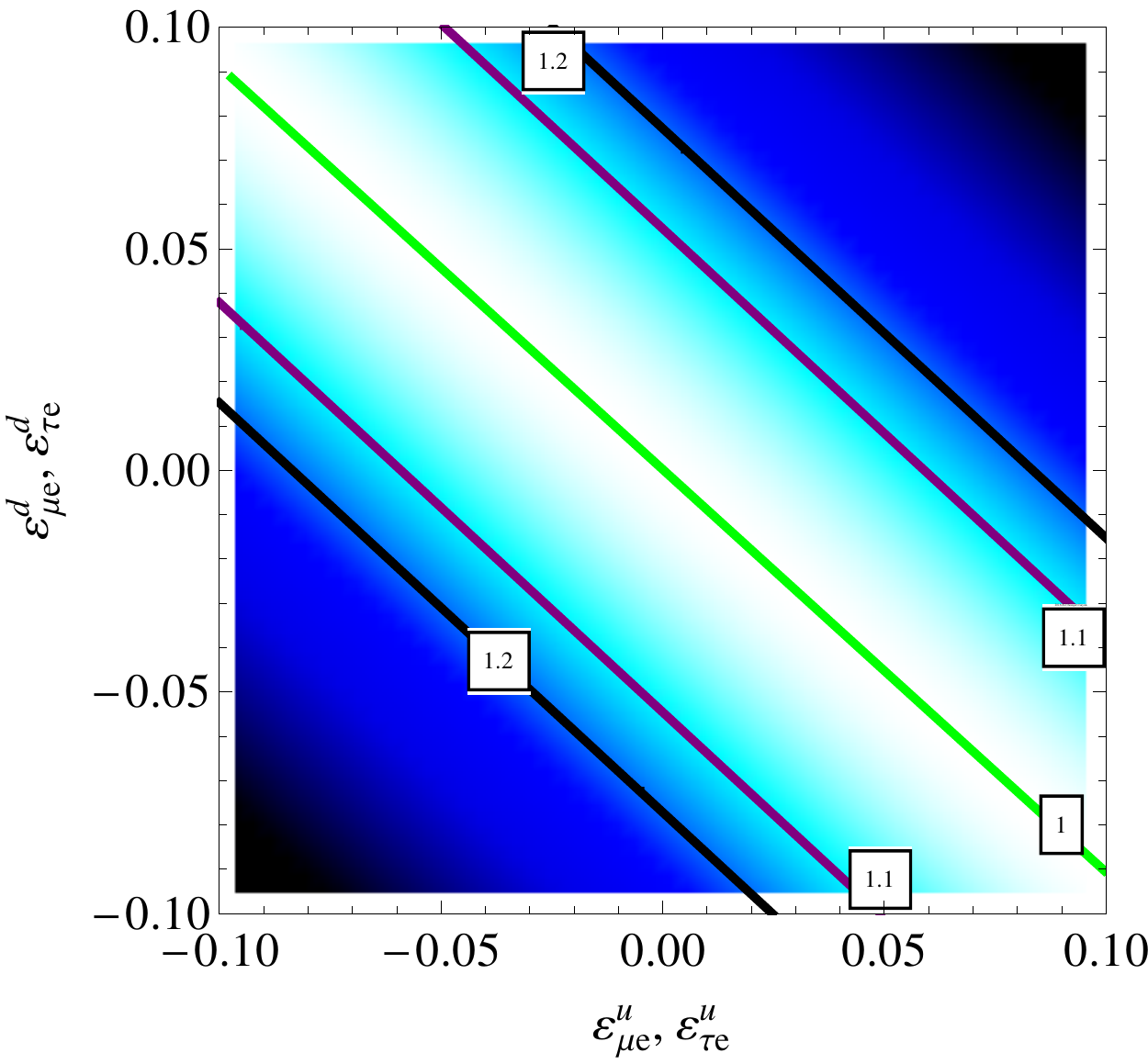}

\includegraphics[width=6cm]{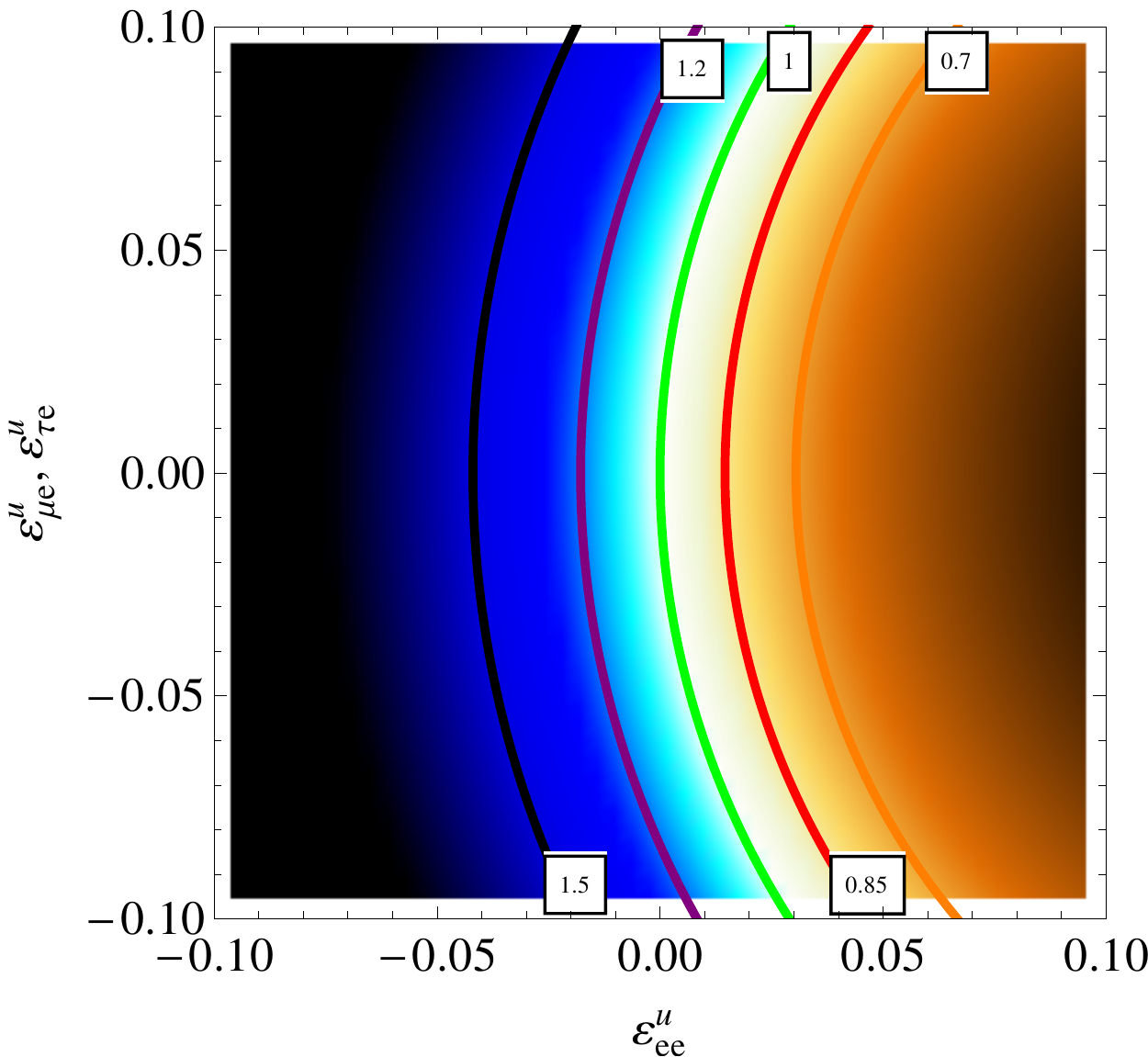}\includegraphics[width=6cm]{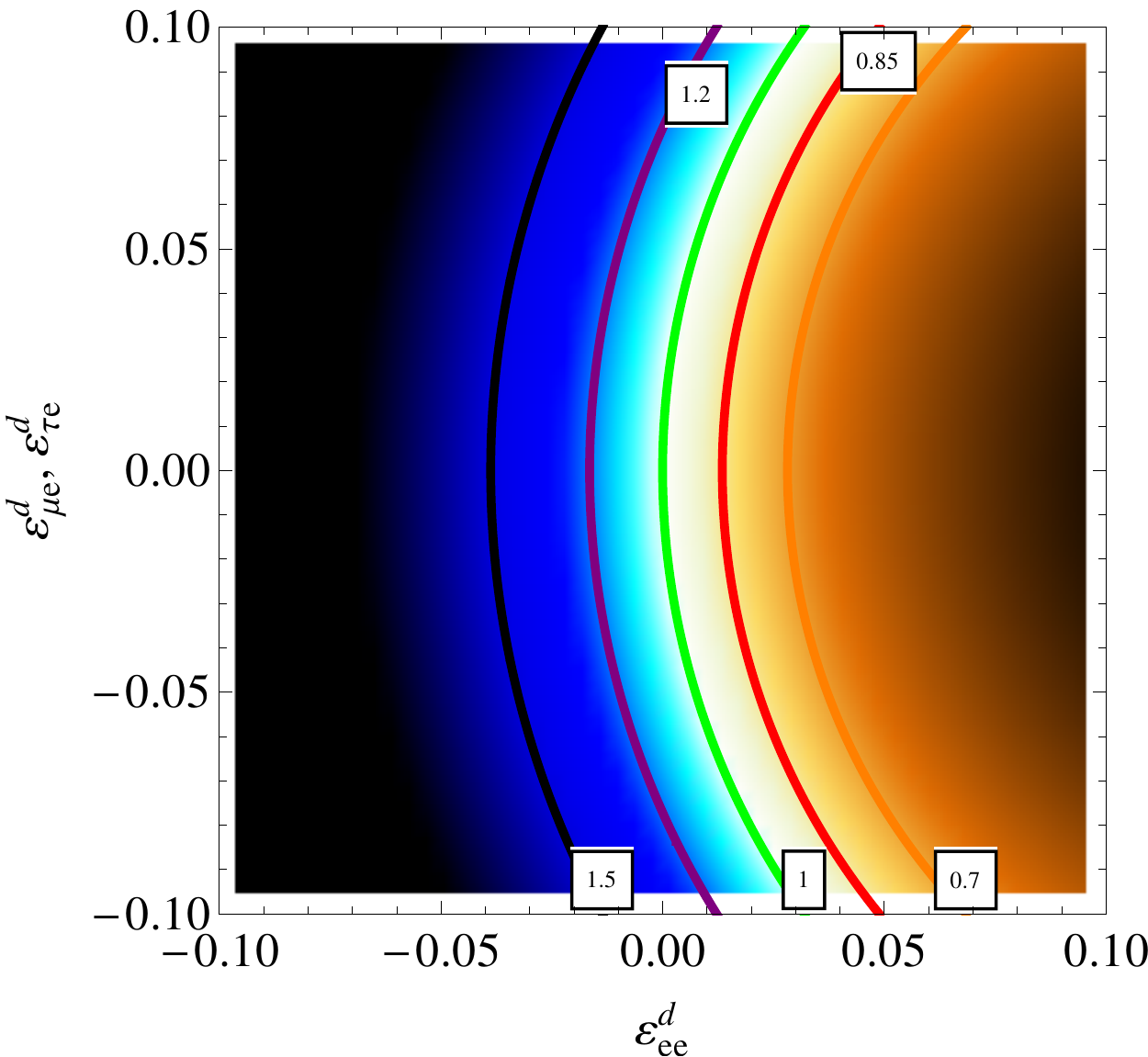}

\includegraphics[width=6cm]{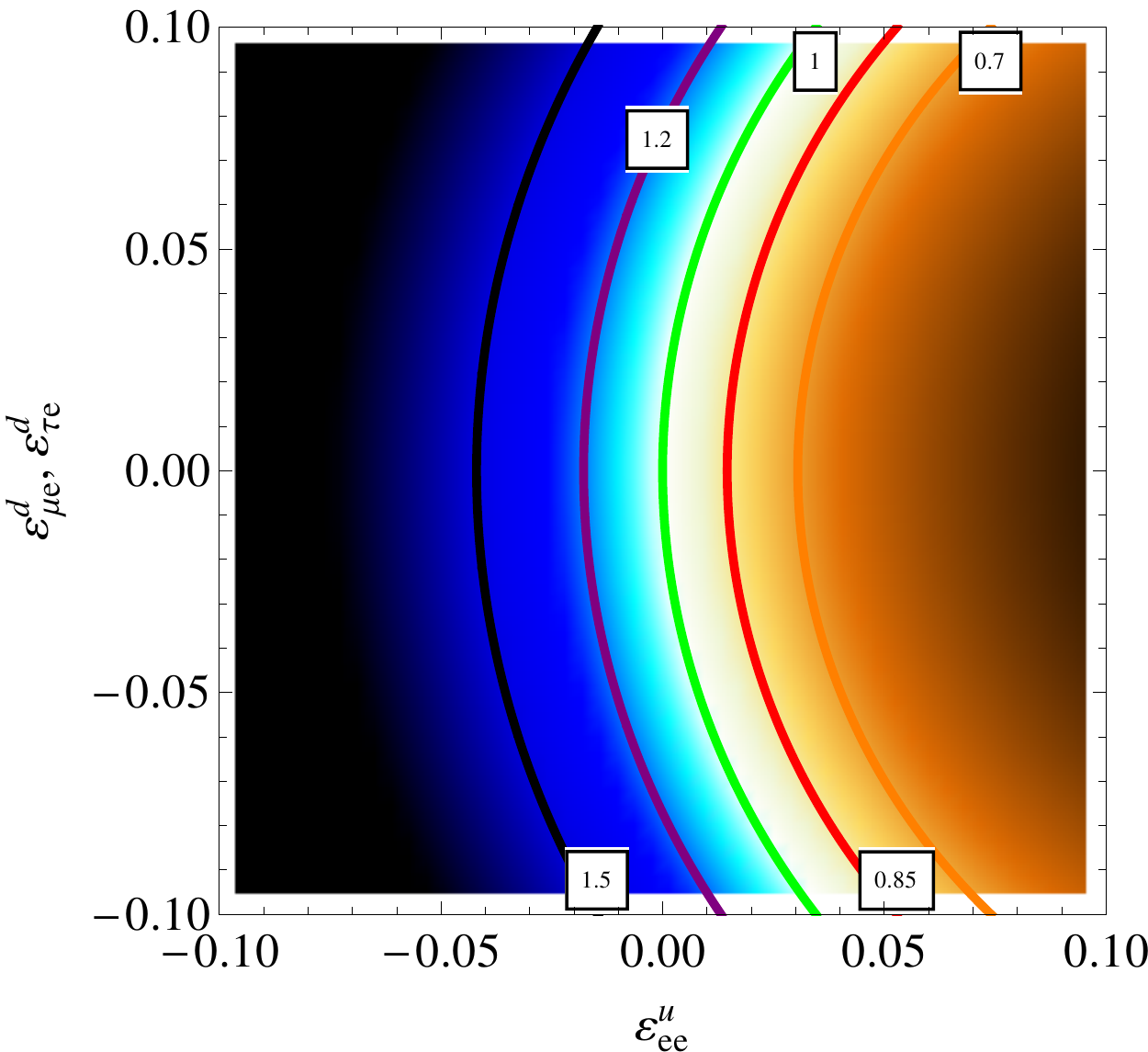}\includegraphics[width=6cm]{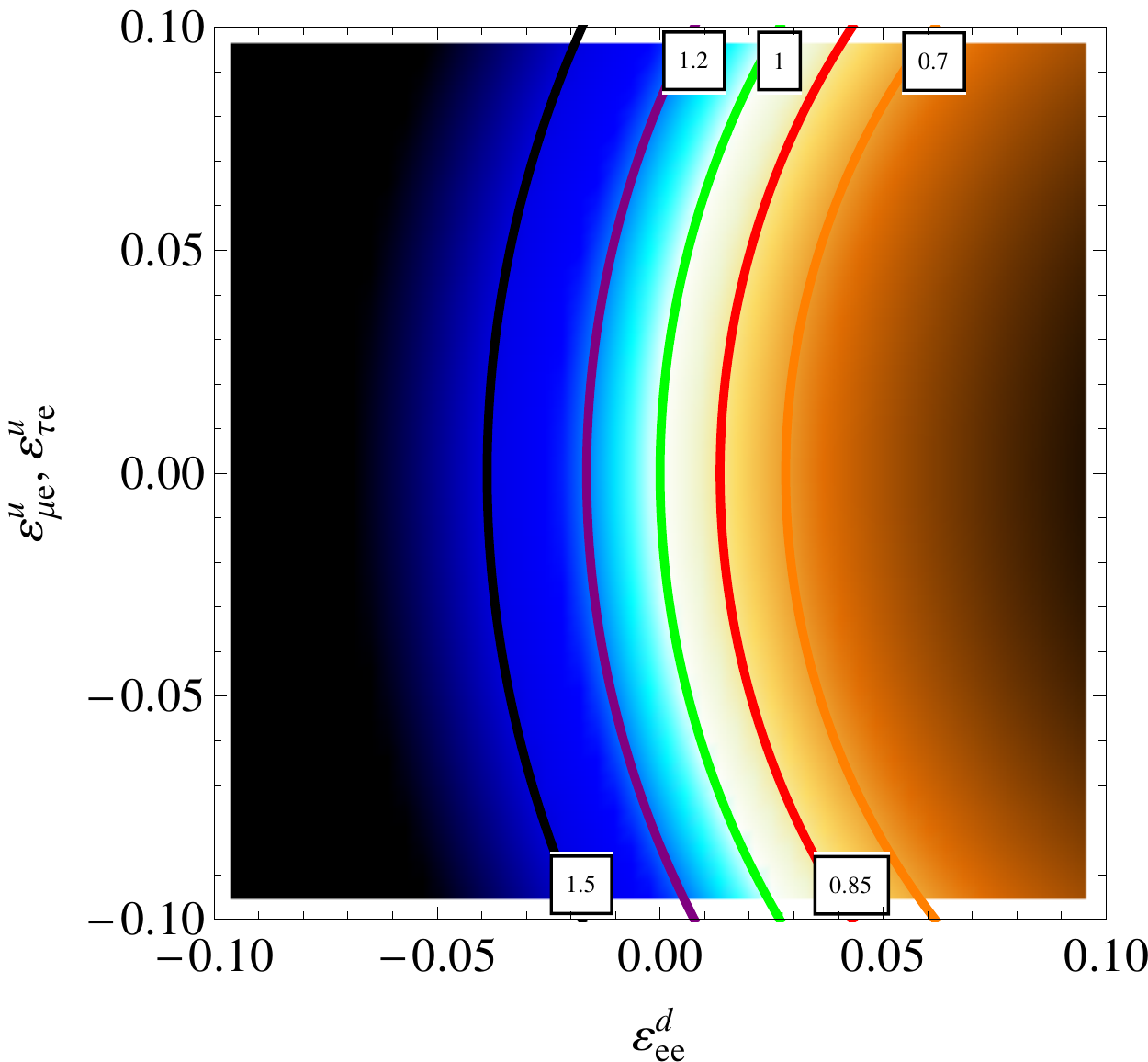}

\end{minipage}\begin{minipage}{3cm}

\includegraphics[width=1.5cm]{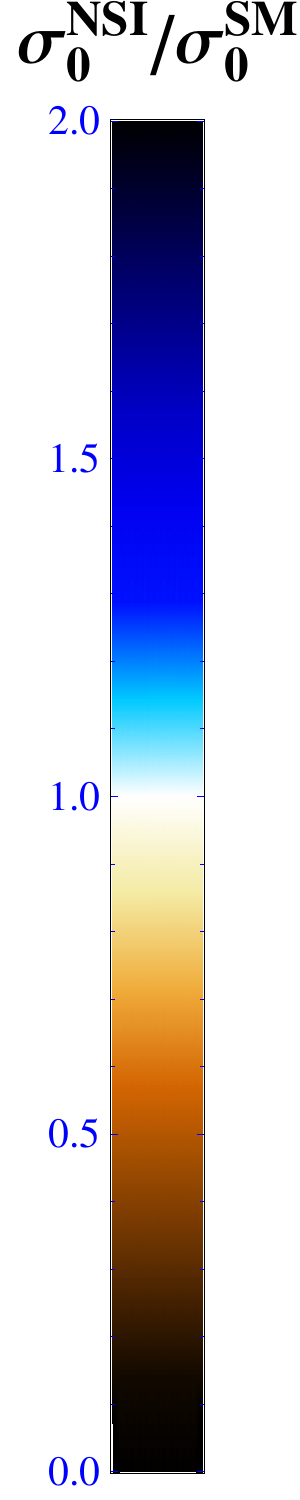}

\end{minipage}%

\caption{\label{fig:NSI-dep} The effect of NSI parameters on the cross section
ratio $\sigma_{0}^{{\rm NSI}}/\sigma_{0}^{{\rm SM}}$.
The lower four plots are similar, because
a large nucleus is almost symmetric with respect to
$u\leftrightarrow d$. }
\end{figure}

To understand how NSI affect C$\nu N$S, we study
the dependence of $\sigma_{0}^{{\rm NSI}}$ on the six parameters
in Eq.\ (\ref{eq:coh-64}) with several plots in Fig.\ \ref{fig:NSI-dep}.
Each plot displays the ratio $\sigma_{0}^{{\rm NSI}}/\sigma_{0}^{{\rm SM}}$
as a function of two $\varepsilon$ in Eq.\ (\ref{eq:coh-64}), while
the other four $\varepsilon$ are set to zero. Note that Eq.\ (\ref{eq:coh-63})
is symmetric under exchange of $\mu$ and $\tau$, thus we combine
plots for $\varepsilon_{\mu e}^{q}$ and $\varepsilon_{\tau e}^{q}$
since for coherent $\nu-N$ scattering they have the same effect.

From the top two panels in Fig.\ \ref{fig:NSI-dep} one can see that
there is one direction (the green line)
in which $\sigma_{0}^{{\rm NSI}}/\sigma_{0}^{{\rm SM}}$
is always equal to $1$, approximately at
$\varepsilon_{\alpha e}^{u}\approx-\varepsilon_{\alpha e}^{d}$.
Under the approximation that $N/Z\approx1$ one can immediately derive
this relation from Eq.\ (\ref{eq:coh-63}). It implies that C$\nu N$S
does not have any sensitivity on NSI parameters
along this direction, which has already been discussed
in Refs.\ \cite{Barranco:2005yy,Scholberg:2005qs}.
 In the other panels, the direction with $\sigma_{0}^{{\rm NSI}}/\sigma_{0}^{{\rm SM}}=1$
also exists but in the form of a curve rather than a straight line.
Therefore, degeneracies are present, which in case the NSI actually exist would
need to be broken by other experiments, most notably neutrino oscillation
experiments.


Fig.\ \ref{fig:NSI-dep} also shows that the ratio $\sigma_{0}^{{\rm NSI}}/\sigma_{0}^{{\rm SM}}$ could significantly deviate from $1$. Even for small
values of $\varepsilon$ in the range $(-0.1,\thinspace0.1)$,
$\sigma_{0}^{{\rm NSI}}$ could vanish
($\sigma_{0}^{{\rm NSI}}/\sigma_{0}^{{\rm SM}}=0$) or rise to twice
the SM value ($\sigma_{0}^{{\rm NSI}}/\sigma_{0}^{{\rm SM}}=2$).
Therefore once coherent $\nu-N$ scattering is observed, it will provide
a significant constraint on NSI parameters. Besides, among the six
plots in Fig.\ \ref{fig:NSI-dep}, only the top right one does not
include $\sigma_{0}^{{\rm NSI}}/\sigma_{0}^{{\rm SM}}<1$, which implies
that if the measured cross section is lower than the SM value, then
$\varepsilon_{ee}^{u}$ or $\varepsilon_{ee}^{d}$ have to be non-zero
in order to explain the deficit by NSI.
A statistical analysis of the sensitivity on NSI will be performed in
section \ref{sec:fit}.

\section{\label{sec:Exotic}Exotic Neutral Currents in
Coherent $\nu-N$ Scattering}

Apart from the NSI which only couple neutrinos to quarks in (chiral)
vector form, more ``exotic'' new interactions could be present.
There are five types of possible interactions, scalar ($S$),
pseudo-scalar ($P$), vector ($V$), axial-vector ($A$), and tensor
($T$) interactions:\footnote{To make the following calculation more compact, we assume that the
SM neutral current interaction is included in Eq.\ (\ref{eq:coh-35})
rather than adding Eq.\ (\ref{eq:coh-35}) to the SM Lagrangian. As
a consequence, in the SM $C_{a}^{(q)}$ and $\overline{D}_{a}^{(q)}$ are non-zero. }
\begin{equation}
{\cal L}\supset\frac{G_{F}}{\sqrt{2}}\sum_{a=S,P,V,A,T}\overline{\nu}\,\Gamma^{a}\nu\left[\overline{q}\Gamma^{a}(C_{a}^{(q)}+\overline{D}_{a}^{(q)}i\gamma^{5})q\right],\label{eq:coh-35}
\end{equation}
where $q$ stands for $u$ and $d$ quarks and
\begin{equation}
\Gamma^{a}=\{I,i\gamma^{5},\gamma^{\mu},\gamma^{\mu}\gamma^{5},\sigma^{\mu\nu}\equiv\frac{i}{2}[\gamma^{\mu},\gamma^{\nu}]\}.\label{eq:coh-42}
\end{equation}
In analogy to Eq.\ (\ref{eq:NSITeV}), the $C_{a}^{(q)}$
and $\overline{D}_{a}^{(q)}$ are expected to be of order
$(g_X^2/g^2)  \, (M_W^2 /M_X^2) $, with new exchange particles $M_X$ and
coupling constants $g_X$.
The coefficients $C_{a}^{(q)}$ and $\overline{D}_{a}^{(q)}$ in Eq.\,(\ref{eq:coh-35})
are dimensionless and in principle can be complex numbers. However
if the interaction term is not self-conjugate, it would be added by
its complex conjugate, which is proportional to $\overline{\nu}\,\Gamma^{a}\nu\left[\overline{q}\Gamma^{a}(C_{a}^{(q)*}+\overline{D}_{a}^{(q)*}i\gamma^{5})q\right]$
for $a=S,\thinspace P,\thinspace T$ and $\overline{\nu}\,\Gamma^{a}\nu\left[\overline{q}\Gamma^{a}(C_{a}^{(q)*}-\overline{D}_{a}^{(q)*}i\gamma^{5})q\right]$
for $a=V,\thinspace A$. Since $C_{a}^{(q)*}+C_{a}^{(q)}$, $\overline{D}_{a}^{(q)}+\overline{D}_{a}^{(q)*}$
and $i(\overline{D}_{a}^{(q)}-\overline{D}_{a}^{(q)*})$ are real
numbers, without loss of generality we can take $C_{a}^{(q)}$ and
\begin{equation}
D_{a}^{(q)}\equiv\begin{cases}
\overline{D}_{a}^{(q)} & (a=S,\thinspace P,\thinspace T)\\
i\overline{D}_{a}^{(q)} & (a=V,\thinspace A)
\end{cases}\label{eq:coh-43}
\end{equation}
as real numbers. We will assume for simplicity that $C_{a}^{(u)} = C_{a}^{(d)}$
and $\overline{D}_{a}^{(u)} = \overline{D}_{a}^{(d)}$. This still leaves us with
10 free parameters.

A subtle issue related to $\sigma^{\mu\nu}$ and $\sigma^{\mu\nu}\gamma^{5}$
should be clarified here. When the tensor $\overline{\nu}\sigma^{\mu\nu}\nu$
is coupled to $\overline{q}\sigma^{\mu\nu}q$, there are two possibilities,
$\overline{\nu}\sigma^{\mu\nu}\nu\overline{q}\sigma_{\mu\nu}q$ and
$\epsilon^{\mu\nu\rho\sigma}\overline{\nu}\sigma_{\mu\nu}\nu\overline{q}\sigma_{\rho\sigma}q$.
On the other hand, there could be new interactions such as $\overline{\nu}\sigma^{\mu\nu}\gamma^{5}\nu\overline{q}\sigma_{\mu\nu}q$
and $\overline{\nu}\sigma^{\mu\nu}\gamma^{5}\nu\overline{q}\sigma_{\mu\nu}\gamma^{5}q$,
which seem not to be included in Eq.\ (\ref{eq:coh-35}). But due
to the identity
\begin{equation}
\sigma^{\mu\nu}i\gamma^{5}=-\frac{1}{2}\sigma_{\rho\sigma}\epsilon^{\mu\nu\rho\sigma}\label{eq:coh-44}
\end{equation}
all these new possibilities can be transformed into the tensor form appearing
in Eq.\ (\ref{eq:coh-35}):
\begin{equation}
\overline{\nu}\sigma^{\mu\nu}\gamma^{5}\nu\overline{q}\sigma_{\mu\nu}q=\frac{i}{2}\epsilon^{\mu\nu\rho\sigma}\overline{\nu}\sigma_{\rho\sigma}\nu\overline{q}\sigma_{\mu\nu}\gamma^{5}q=\overline{\nu}\sigma^{\mu\nu}\nu\overline{q}\sigma_{\mu\nu}\gamma^{5}q\,.\label{eq:coh-45}
\end{equation}

Since the coherent nature of the scattering
requires low energy, we can treat the nucleus
in the coherent scattering as a point-like particle. Depending on
the spin of the nucleus, it can be described by a scalar field, a
Dirac field or even higher spin fields. As we have shown in Eq.\
(\ref{eq:coh-41}), for low energy scattering the difference of treating
the nucleus as a spin-0 or spin-1/2 particle is negligible, and in fact
identical to order $(T/E_{\nu})^{2}$. In the following
calculation we will treat the nucleus as a spin-1/2 particle since
 for automatic calculation implemented by packages (we use both
FeynCalc \cite{Shtabovenko:2016sxi,Mertig:1990an} and Package-X
\cite{Patel:2015tea}) it is technically simpler
than the scalar treatment.
Consequently, the effective Lagrangian of neutrino-nucleus interactions has
the same form as Eq.\ (\ref{eq:coh-35}) with $q$ replaced by the
Dirac field $\psi_{N}$ of the nucleus, i.e.\
\begin{equation}
{\cal L}\supset\frac{G_{F}}{\sqrt{2}}\sum_{a=S,P,V,A,T}\overline{\nu}\Gamma^{a}\nu\left[\overline{\psi_{N}}\Gamma^{a}(C_{a}+\overline{D}_{a}i\gamma^{5})\psi_{N}\right].\label{eq:coh-94}
\end{equation}
Note that to define the effective couplings of $\psi_{N}$ to $\nu$,
here we use $(C_{a},\thinspace\overline{D}_{a})$ which should be
related to the more fundamental couplings $(C_{a}^{(q)},\thinspace\overline{D}_{a}^{(q)})$.
Since the relations are lengthy and also involve form factors,
we present them in Appendix \ref{sec:formfactor}. From now on, we will
consider $C_{a}$ and $\overline{D}_{a}$ as parameters of interest, and will
present results in terms of those.
We are not aware of literature limits on the parameters, which would
have been obtained from past neutrino-nucleon scattering experiments. Since the
event numbers in our benchmark experiment are much larger than in such
experiments, the sensitivities we will derive later would surely be
orders of magnitude better.

From Eq.\ (\ref{eq:coh-94}), we can write down the scattering amplitude,
\begin{eqnarray}
i{\cal M}^{s'sr'r} & = & -i\frac{G_{F}}{\sqrt{2}}\overline{v}^{s}(p_{1})P_{R}\Gamma^{a}v^{s}(k_{1})\overline{u}^{r'}(k_{2})\Gamma^{a}(C_{a}+\overline{D}_{a}i\gamma^{5})u^{r}(p_{2}) \, . \label{eq:coh-57}
\end{eqnarray}
Note that for general interactions, the coherent cross sections of $\nu N$
and $\overline{\nu}N$ are different {[}in the SM coherent
$\nu N$ and $\overline{\nu}N$ cross sections are the same
due to the approximate parity symmetry in nuclei, see comments after
Eq.\ (\ref{eq:coh-39-1}){]}.
Since we are studying the coherent scattering of reactor
neutrinos, only right-handed antineutrinos are considered. Therefore
we have attached a $P_{R}=(1+\gamma^{5})/2$ projection to the initial
neutrino state $\overline{v}^{s}(p_{1})$, so that the trace technology
applies,
\begin{equation}
|{\cal M}|^{2}=\sum_{ss'}\frac{1}{2}\sum_{rr'}|{\cal M}^{s'sr'r}|^{2}\,.\label{eq:coh-58}
\end{equation}
The result is given by
\begin{eqnarray}
\frac{d\sigma}{dT} & = & \frac{G_{F}{}^{2}M}{4\pi}N^{2}\left[\xi_{S}^{2}\frac{MT}{2E_{\nu}{}^{2}}\right.\nonumber \\
 &  & +\xi_{V}^{2}\left(1-\frac{T}{T_{{\rm max}}}\right)-2\xi_{V}\xi_{A}\frac{T}{E_{\nu}}+\xi_{A}^{2}\left(1-\frac{T}{T_{{\rm max}}}+\frac{MT}{E_{\nu}{}^{2}}\right)\nonumber \\
 &  & +\xi_{T}^{2}\left(1-\frac{T}{T_{{\rm max}}}+\frac{MT}{4E_{\nu}{}^{2}}\right)\nonumber \\
 &  & \left.-R\frac{T}{E_{\nu}}+{\cal O}\left(\frac{T^{2}}{E_{\nu}^{2}}\right)\right],\label{eq:coh-55}
\end{eqnarray}
where
\begin{equation}
\xi_{S}^{2}=\frac{1}{N^{2}}(C_{S}^{2}+D_{P}^{2}),\thinspace\xi_{T}^{2}=\frac{8}{N^{2}}\left(C_{T}^{2}+D_{T}^{2}\right),\thinspace\xi_{V}=\frac{1}{N}(C_{V}-D_{A}),\thinspace\xi_{A}=\frac{1}{N}(C_{A}-D_{V})\label{eq:coh-59}
\end{equation}
and
\begin{equation}
R\equiv\frac{2}{N^{2}}(C_{P}C_{T}-C_{S}C_{T}+D_{T}D_{P}-D_{T}D_{S})\,.\label{eq:coh-60}
\end{equation}
As we can see, the cross section only depends on 5 parameters,
$\overrightarrow{\xi}\equiv(\xi_{S},\thinspace\xi_{V},\thinspace\xi_{A},\thinspace\xi_{T},\thinspace R)$, compared to the
10 parameters in Eq.\ (\ref{eq:coh-94}).

The first three lines
of Eq.\ (\ref{eq:coh-55}) come from  scalar and pseudo-scalar, vector
and axial vector, and tensor interactions respectively while the $R$
term is an interference term of the (pseudo-) scalar and tensor interactions.
Despite that $\xi_{S}$ contains both scalar and pseudo-scalar contributions,
for simplicity we will refer to  $\xi_{S}^{2}$ as the scalar
interaction of neutrinos with nuclei. In the same way, though the
vector couplings $(C_{V},\thinspace D_{V})$ and the axial vector
couplings $(C_{A},\thinspace D_{A})$ all appear
in $(\xi_{V},\thinspace\xi_{A})$,
we still call $\xi_{V}^{2}$ and $\xi_{A}^{2}$ the vector
and axial interactions, respectively.

Comparing Eq.\ (\ref{eq:coh-55}) to Eq.\ (\ref{eq:coh-34}), we obtain
the SM values of these parameters,
\begin{equation}
\overrightarrow{\xi}{}_{{\rm SM}}\equiv(0,\thinspace1-(1-4s_{W}^{2})Z/N,\thinspace0,\thinspace0,\thinspace0)\approx(0,\thinspace0.962,\thinspace0,\thinspace0,\thinspace0)\,,\label{eq:coh-61}
\end{equation}
where the number $0.962$ is computed for Germanium, i.e.\
by taking $N=40.6$, $Z=32$ and $s_{W}^{2}=0.238$.

\begin{figure}
\centering

\includegraphics[width=8cm]{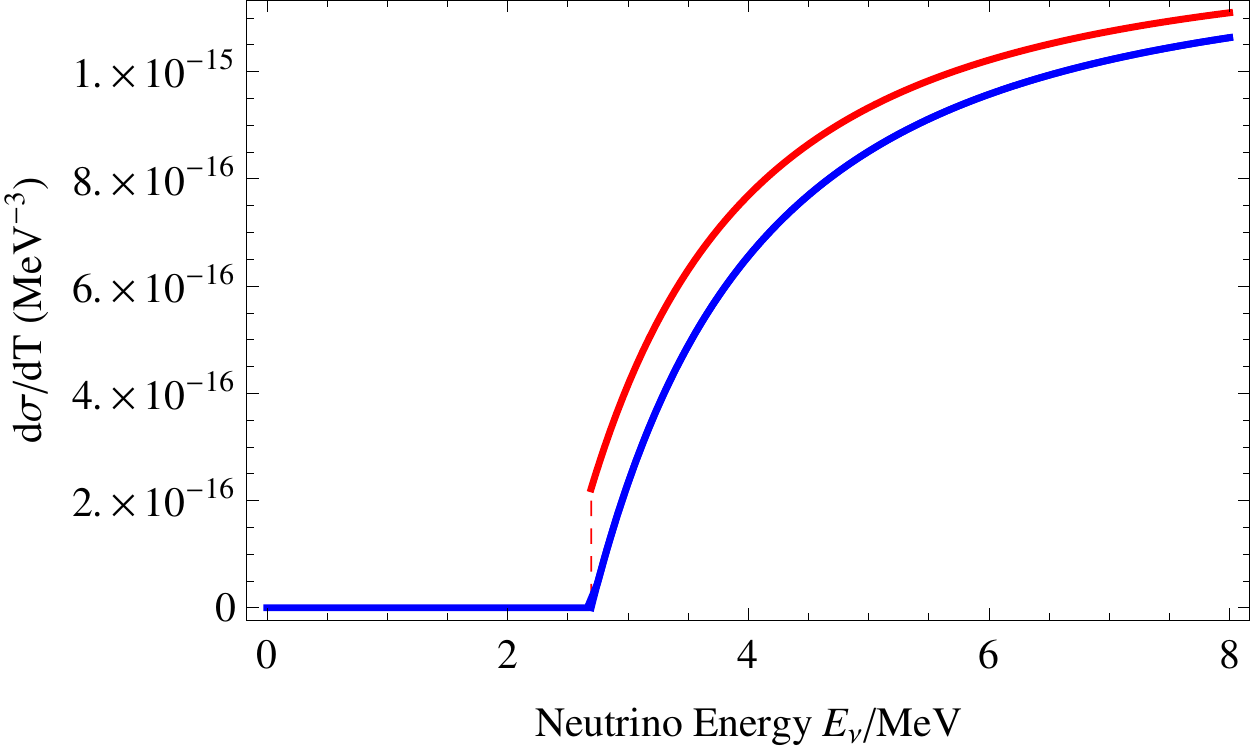}

\caption{\label{fig:non0thre}Effect of exotic neutral currents near the
threshold. We plot $\frac{d\sigma}{dT}$ as a function
of $E_{\nu}$ according to Eq.\ (\ref{eq:coh-55}) with fixed threshold
$T_{\rm th} = 0.2$ keV, corresponding to $E_{\nu}\gs 2.7$
MeV for neutrinos. At $E_{\nu} = 2.7$ MeV, $T = T_{\rm max}$ and the
cross section (\ref{eq:coh-34}) in the SM vanishes.
For exotic neutral currents the cross section (\ref{eq:coh-55}) does
not vanish for $T = T_{\rm max}$.
The parameters are $\protect\overrightarrow{\xi}=\protect\overrightarrow{\xi}{}_{{\rm SM}}$
for the blue curve and $\protect\overrightarrow{\xi}=(0.4,\thinspace1,\thinspace0.1,\thinspace0.1,\thinspace0)$
for the red curve.}
\end{figure}

There are some noteworthy comments to make from Eq.\ (\ref{eq:coh-55}):
\begin{itemize}
\item There is no interference term of (axial) vector interactions with
other interactions. But the vector interaction interferes with the
axial interaction.
\item The energy dependence of the $\xi_{V}^{2}$ term is the same as that
in the SM {[}cf.\ Eqs.\ (\ref{eq:coh-34}) and (\ref{eq:coh-29-1}){]}.
Hence, new vector interactions  will not distort the recoil energy
spectrum.
\item The other terms (i.e.\ scalar, axial, tensor interaction terms and
two interference terms) have different energy dependence. If
any distortion on the recoil energy spectrum would be observed, then
these new interactions could be the explanation.
\item For vector interactions,
$\frac{d\sigma}{dT}$ is zero at $T_{\max}(E_{\nu})$
{[}defined in Eq.\ (\ref{eq:coh-29-1}){]} but it could be non-zero
if other types of interactions exist. This is shown in Fig.\
\ref{fig:non0thre} where at the threshold the cross section is seen to be
zero (blue curve) for the SM but non-zero (red curve) if other
types of exotic neutral currents exist.
\item Introducing exotic neutral currents (except for vector interactions)
can not reduce the cross section since the sum of the other terms
besides the $\xi_{V}^{2}$ term in Eq.\ (\ref{eq:coh-55}) is  always
above zero. So if the observed events are less than the expectation
from the SM, one should consider modifications only limited to the
vector sector rather than introducing scalar or tensor interactions.
\end{itemize}

\begin{figure}
\centering

\includegraphics[width=5.5cm,height=4cm]{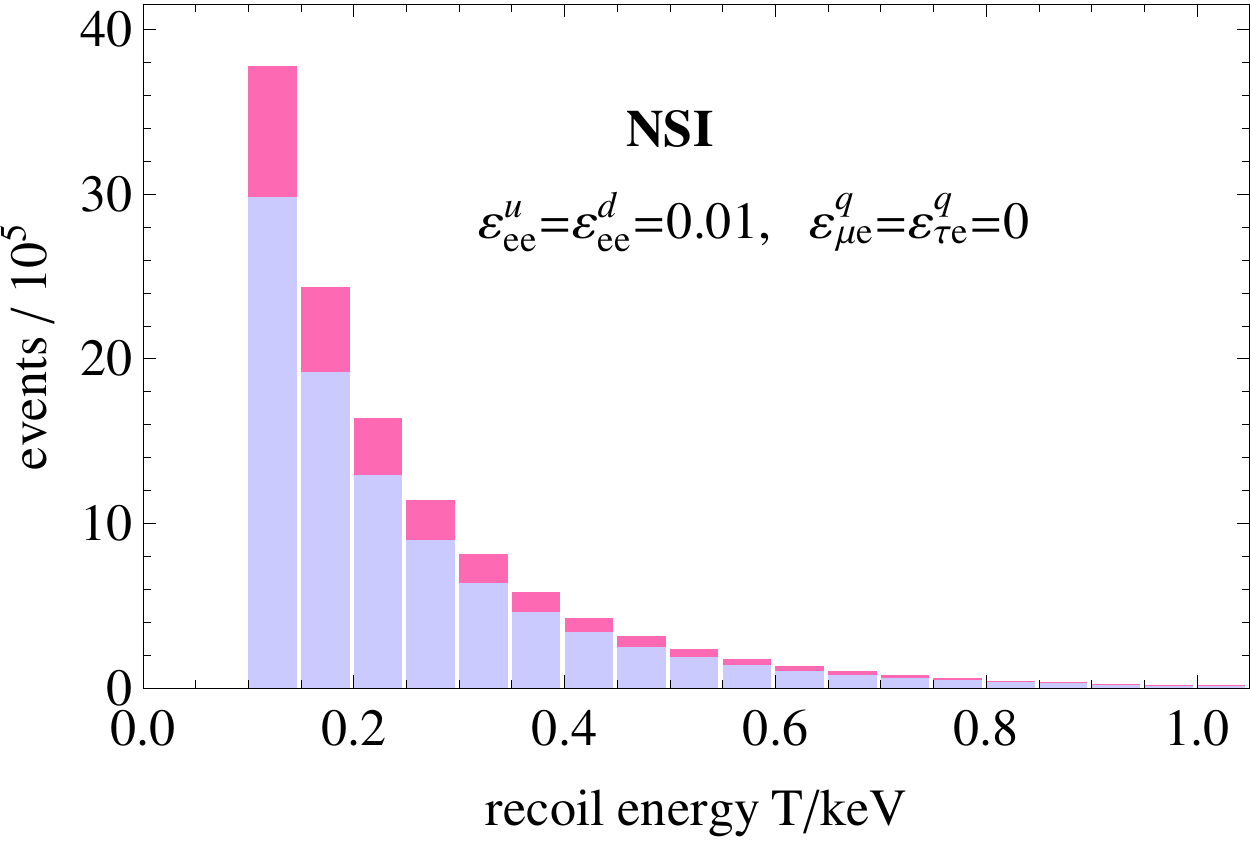}\,\,\includegraphics[width=5.5cm,,height=4cm]{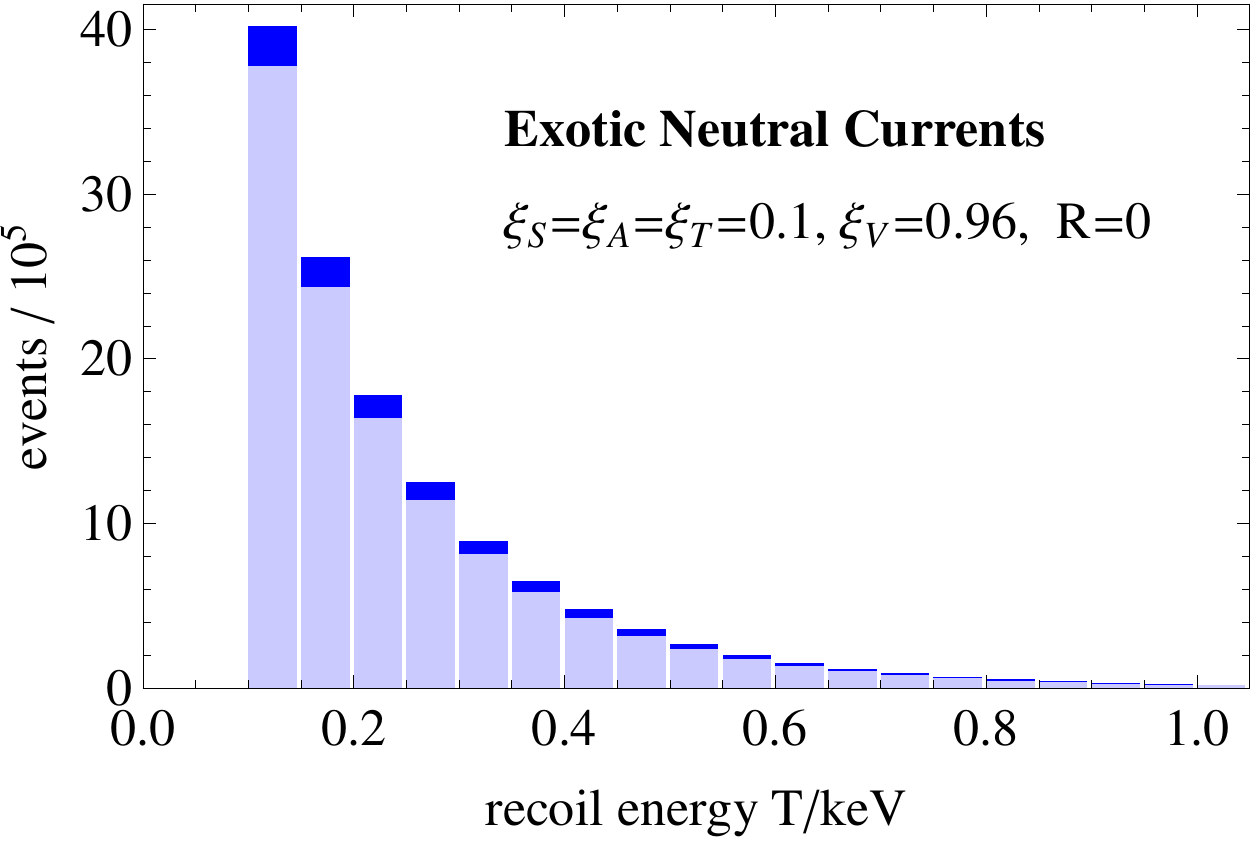}\,\,\includegraphics[width=5.5cm,,height=4cm]{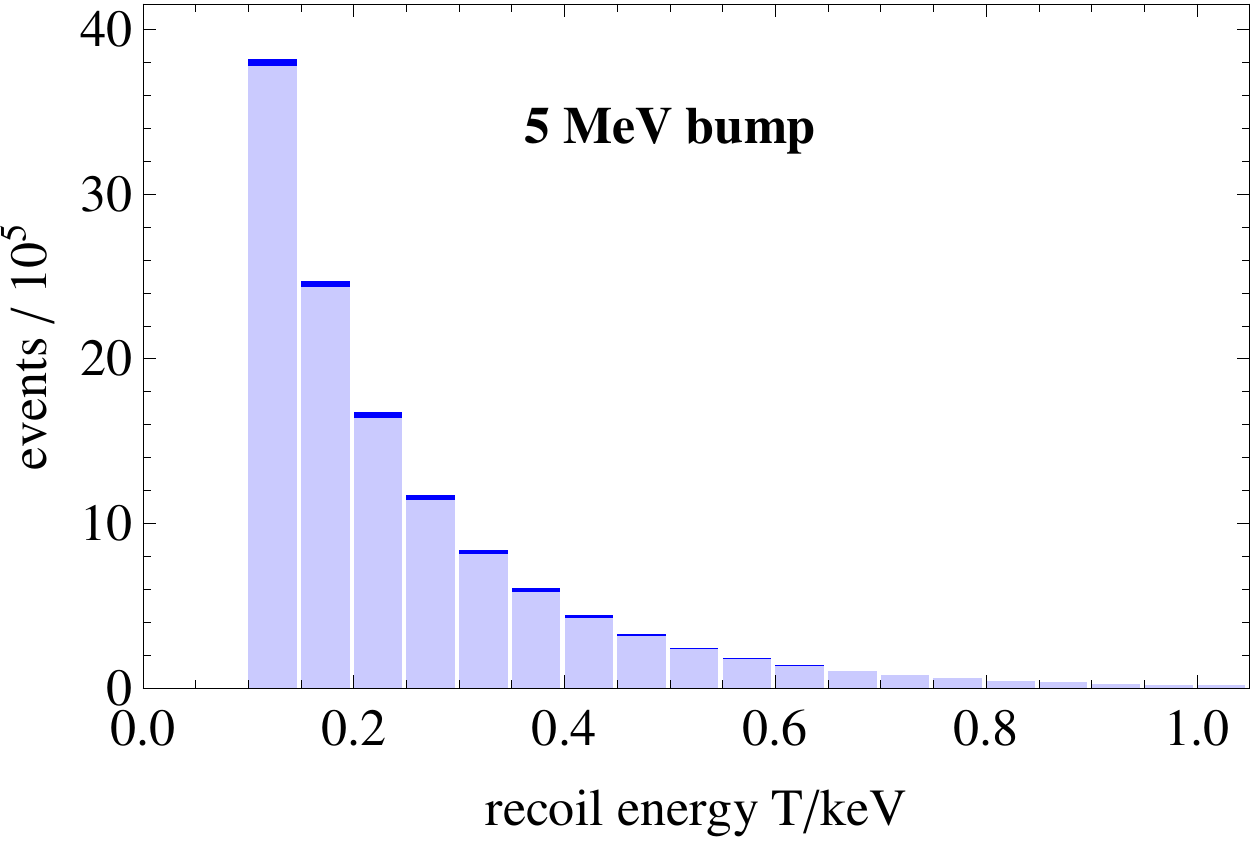}

\caption{\label{fig:events-3case} Event excess/deficit due to several possible
modifications. The pink color is for deficit and dark blue for excess.
}
\end{figure}

\begin{figure}[t]
\centering

\includegraphics[width=9cm]{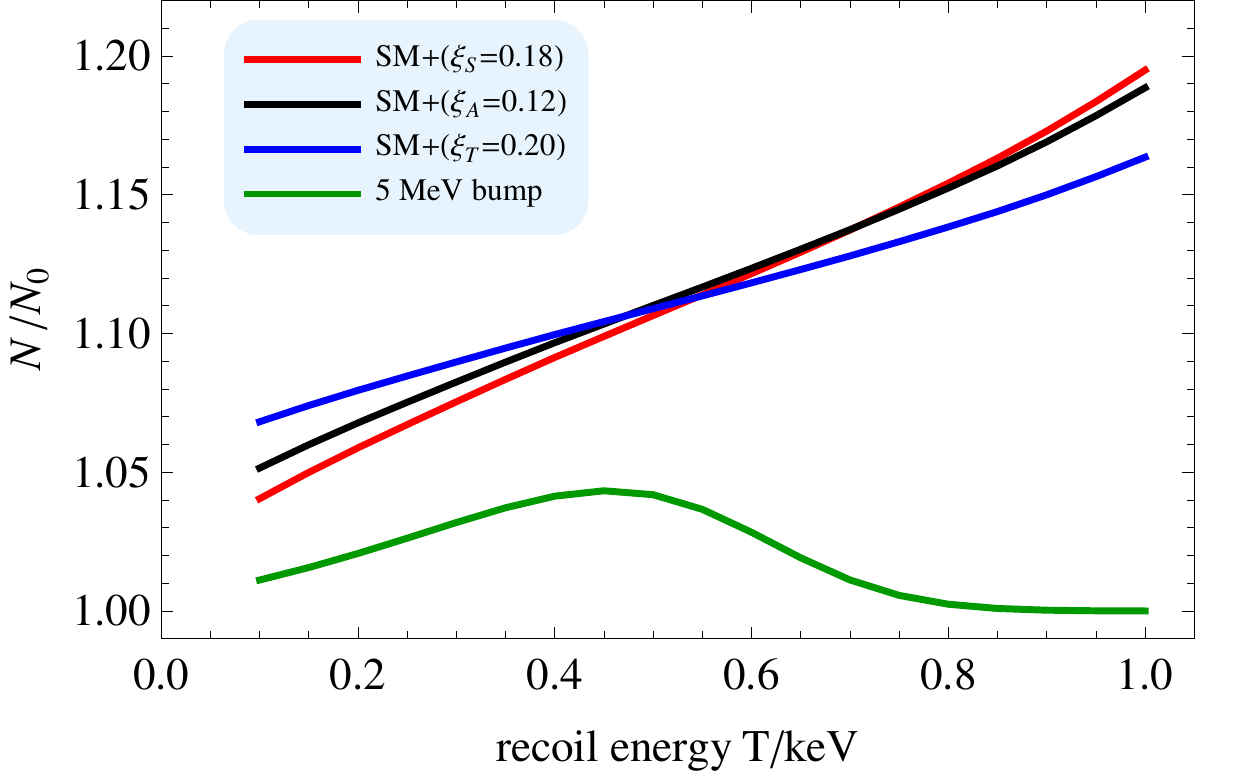}

\caption{\label{fig:distortion} Distortion of the spectrum due to exotic neutral
currents and the 5 MeV bump. The red, black and blue
curves correspond to scalar, axial vector and tensor interactions
in addition to the SM. The green curve is produced by including the
5 MeV bump in the neutrino flux, taken from Ref.\ \cite{Huber:2016xis}. }
\end{figure}

The various modifications we have discussed so far (NSI, exotic neutral
currents and the 5 MeV bump) can influence the event numbers.
In Fig.\ \ref{fig:events-3case} we illustrate this for
three examples.
A feature of NSI is that they could result in a significant
deficit (excess is also possible) of the event number, whereas exotic
neutral currents only lead to an excess if $\xi_{V}$ is fixed at
its SM value. In principle exotic neutral currents could also lead
to a deficit by lowering $\xi_{V}$, but this is indistinguishable from the
NSI case. The 5 MeV bump in the neutrino
flux also leads to an excess, but is not very significant.
Here we take the size of the 5 MeV bump from a recent fit in Ref.\
\cite{Huber:2016xis} (given by its Fig.\ 2). The excess in the 0.10-0.15
keV bin is only about 1\%, which can be easily hidden in the systematic
uncertainties. As mentioned before,
since other experiments will collect with different reactor types
a large amount of event numbers around the 5 MeV bump,
it is very likely that before a highly sensitive Ge detector with
very small systematic uncertainties is running, the 5 MeV bump problem
will be solved (both in theory and experiment).

Another important difference is that the above three cases have very
different effects on the distortion of the spectrum. NSI will not
lead to any distortion at all since it only changes the overall factor
in the differential cross section while the other two cases, exotic
neutral currents and the 5 MeV bump, lead to different distortions.
 In Fig.\ \ref{fig:distortion} we show variations of the event
ratio $N/N_{0}$  as a function of $T$ in several situations, where $N_{0}$
is the event number expected from for the SM and $N$ includes new
interactions or the 5 MeV bump. For exotic neutral currents, we plot
three examples to illustrate the effects from scalar, axial vector
and tensor interactions with $\xi_{S}=0.18$, $\xi_{A}=0.12$ and
$\xi_{T}=0.20$ respectively. All the other parameters, if not mentioned,
have been set to the SM values given by Eq.\ (\ref{eq:coh-61}).
As one can see, for exotic neutral currents the ratios increase with
$T$ but the slopes are different. Scalar interactions would produce
the strongest distortion on the spectrum followed by axial vector
and then tensor. The 5 MeV bump also generates an increasing ratio
with respect to $T$ below 0.45 keV. However, the ratio drops down
at higher energies and finally reaches 1. The reason is that neutrinos
at 5 MeV will only contribute to the events below $0.7$ keV
{[}cf.\ Eq.\ (\ref{eq:coh-29-1}){]}. Thus in the range close to but
less than 0.7 keV, the events from the 5 MeV bump should quickly decrease.
If all neutrinos in the bump only had energies exactly at
5 MeV, then the contribution should completely vanish above 0.7 keV.
However, taking the width of the bump into consideration, the actual
limit is a little higher than 0.7 keV.

\section{\label{sec:fit}Sensitivities from a $\chi^{2}$-fit}

In this section, we will adopt $\chi^{2}$-fit to study the sensitivities
of such our assumed future experiment. For convenience, let us state again our
assumed exposure of $5\, {\rm kg \cdot yr \cdot GW \cdot m^{-2}}$ from Eq.\ (\ref{eq:expo}),
corresponding e.g.\ to a 100 kg Germanium detector running for 5 years,
located at a distance of 10 m from a reactor with 1 GW thermal power,
normalized to a total flux of $1.7\times10^{13}\thinspace\text{cm}^{-2}\thinspace{\rm s}^{-1}$.
We will assume different thresholds of $T = 0.1$, $0.2$ or $0.4$ keV, and a constant background of
$3$ cpd = $3$ ${\rm day}^{-1}\thinspace{\rm kg}^{-1}\thinspace{\rm keV}^{-1}$.

\subsection{Statistical Treatment}

Because the event number in each bin is very
large, and thus almost in a Gaussian distribution, we can
take the following $\chi^{2}$-function
\begin{equation}
\chi^{2}(\xi,\thinspace a)=\frac{a^{2}}{\sigma_{a}^{2}}+\sum_{T\thinspace{\rm bins}}\frac{[(1+a)N_{i}(\xi)-N_{i}^{0}]^{2}}{\sigma_{{\rm stat},i}^{2}+\sigma_{{\rm sys},i}^{2}},\label{eq:coh-73}
\end{equation}
where $\xi$ denotes generally the parameters of interest, e.g.\ $\varepsilon_{\alpha\beta}^{q}$
for the NSI case or $(\xi_{S},\thinspace\xi_{V},\thinspace\xi_{A},\thinspace T,\thinspace R)$
for exotic neutral currents. The event numbers in each bin as expected in the SM are denoted as
$N_{i}^{0}$. The statistical uncertainty $\sigma_{{\rm stat},i}$
and the systematic uncertainty $\sigma_{{\rm sys},i}$ of the event
number in the $i$-th bin are given by
\begin{equation}
\sigma_{{\rm stat},i}=\sqrt{N_{i}+N_{{\rm bkg},\thinspace i}}\,,\,\,\sigma_{{\rm sys},i}=\sigma_{f}(N_{i}+N_{{\rm bkg},\thinspace i})\,.\label{eq:coh-76}
\end{equation}
Here the background $N_{{\rm bkg},\thinspace i}$ is set at 3 cpd
(1 cpd = $1$ ${\rm day}^{-1}\thinspace{\rm kg}^{-1}\thinspace{\rm keV}^{-1}$).
We assume that $\sigma_{{\rm sys},i}$ is proportional
to the event number with a coefficient $\sigma_{f}$.
Many systematic uncertainties simply change the total event number
without leading to strong distortions of the spectrum, e.g.\ the uncertainties
from the evaluation of the total flux of neutrinos, nuclear fuel supply,
detection efficiency, fiducial mass of the detector, distance and
geometry corrections, etc.  To describe this part of systematic uncertainties,
we introduce a normalization factor $a$ with a small uncertainty
$\sigma_{a}$, while the other systematic uncertainties remain in
$\sigma_{{\rm sys},i}$. Of course in a more realistic study one should
parametrize specifically the effect of every systematic uncertainty,
some of which can not be described by this approach.
For the
current stage, we simply adopt  Eq.\ (\ref{eq:coh-73}) for our sensitivity
study, which nevertheless should provide realistic results.

It is sometimes useful to know the value of $a$ at the minimum of
$\chi^{2}$ analytically, which is
\begin{equation}
a_{{\rm min}}=\thinspace\frac{\sum_{i}(N_{i}^{0}-N_{i})N_{i}/(\sigma_{{\rm stat},i}^{2}+\sigma_{{\rm sys},i}^{2})}{\sigma_{a}^{-2}+\sum_{i}N_{i}^{2}/(\sigma_{{\rm stat},i}^{2}+\sigma_{{\rm sys},i}^{2})}\,.\label{eq:coh-74}
\end{equation}
One can use Eq.\ (\ref{eq:coh-74}) to marginalize $a$ and obtain
the $\chi^{2}$-function that we are actually interested in,
\begin{equation}
\chi^{2}(\xi)\equiv\chi^{2}(\xi,\thinspace a_{\min})\,.\label{eq:coh-75}
\end{equation}

\begin{figure}[t]
\centering

\includegraphics[width=9cm]{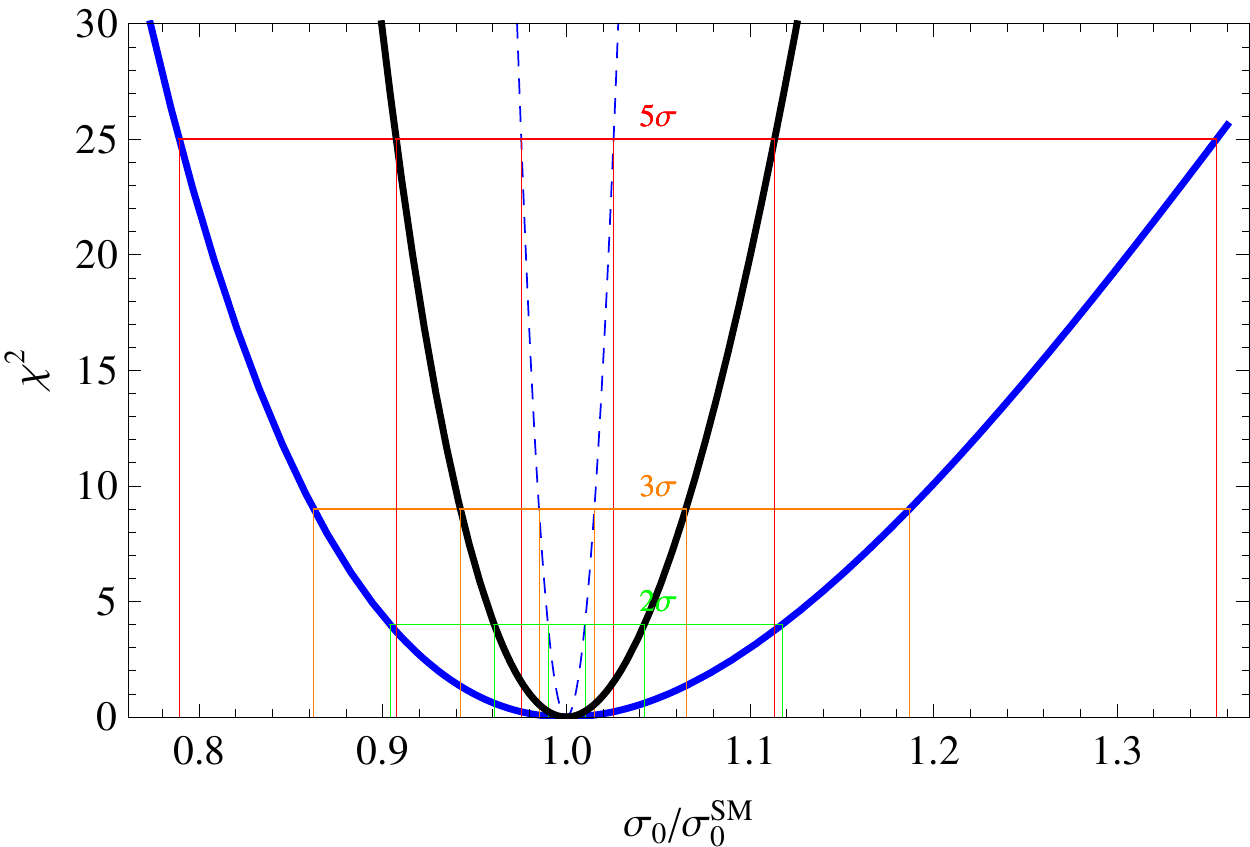}

\caption{\label{fig:sigma0} Sensitivity on the cross section ratio $\sigma_{0}/\sigma_{0}^{\rm SM}$.
The blue solid, black solid and blue dashed curves are generated with
a conservative configuration $(\sigma_{a},\thinspace\sigma_{f},\thinspace T_{{\rm th}})=(5\%,\thinspace3\%,\thinspace0.4\,{\rm keV})$,
an intermediate configuration $(\sigma_{a},\thinspace\sigma_{f},\thinspace T_{{\rm th}})=(2\%,\thinspace1\%,\thinspace0.2\,{\rm keV})$
and an optimistic configuration $(\sigma_{a},\thinspace\sigma_{f},\thinspace T_{{\rm th}})=(0.5\%,\thinspace0.1\%,\thinspace0.1\,{\rm keV})$,
respectively.}
\end{figure}

If coherent $\nu-N$ scattering has been successfully detected, the
first task is to compare the measured total cross section $\sigma_{0}$
with the SM prediction $\sigma_{0}^{{\rm SM}}$
in Eqs.\ (\ref{eq:coh-34})
and (\ref{eq:coh-33}). The ratio $\sigma_{0}/\sigma_{0}^{{\rm SM}}$
indicates any deviation from the SM. One can compute the above $\chi^{2}$-function
to estimate the sensitivity on this ratio ($\xi$ in this case simply
stands for $\sigma_{0}$). The result is shown in Fig.\ \ref{fig:sigma0},
where we have assumed three different
configurations:
\begin{itemize}
\item[(i)] conservative configuration:  $(\sigma_{a},\thinspace\sigma_{f},\thinspace T_{{\rm th}})=(5\%,\thinspace3\%,\thinspace0.4\,{\rm keV})$.
\item[(ii)] intermediate configuration: $(\sigma_{a},\thinspace\sigma_{f},\thinspace T_{{\rm th}})=(2\%,\thinspace1\%,\thinspace0.2\,{\rm keV})$.
\item[(iii)] optimistic configuration: $(\sigma_{a},\thinspace\sigma_{f},\thinspace T_{{\rm th}})=(0.5\%,\thinspace0.1\%,\thinspace0.1\,{\rm keV})$.
\end{itemize}

Even in the conservative configuration, the experiment can measure
$\sigma_{0}/\sigma_{0}^{{\rm SM}}$ with good precision,
$0.862<\sigma_{0}/\sigma_{0}^{{\rm SM}}<1.187$
at 3$\sigma$. In the intermediate case,
$0.942 <\sigma_{0}/\sigma_{0}^{{\rm SM}}<1.065$, while for the
optimistic case $0.985<\sigma_{0}/\sigma_{0}^{\rm SM}<1.015$, all at 3$\sigma$.
As it turns out, the improvement in sensitivity on new physics parameters
between the conservative and intermediate configuration is about a factor of two. Roughly another factor of two can be gained when
going from the intermediate configuration to the somewhat overly optimistic one. The choices we made for the
various configurations should therefore give a feeling on the final sensitivity of such experiments.

\subsection{Low Energy Determination of the Weinberg Angle}
\begin{figure}[t]
\centering

\includegraphics[width=11cm]{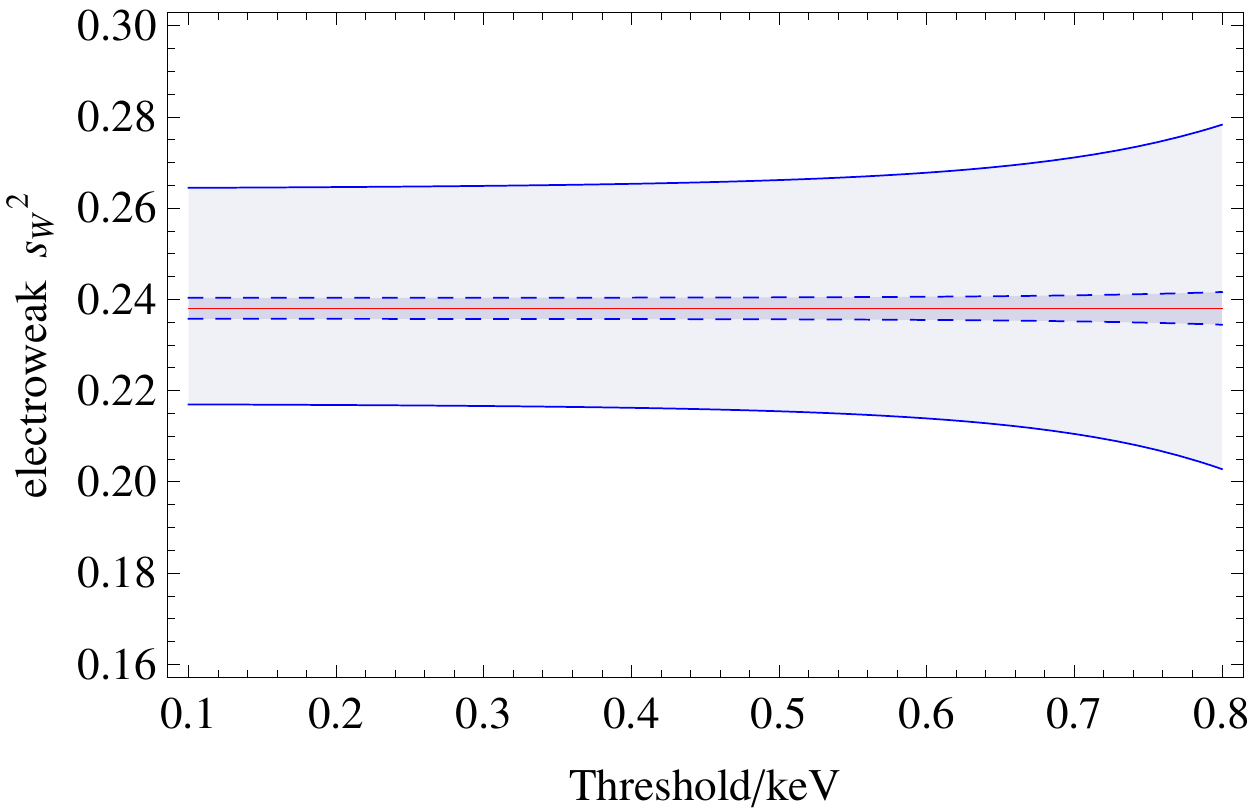}

\caption{\label{fig:swplot} Sensitivity on the electroweak mixing angle $\sin^2 \theta_{W}$.
The central value (red line) is the literature value of $0.238$ and the blue
solid and dashed curves denote 3$\sigma$-bounds in the conservative and optimistic configuration, respectively.}
\end{figure}

The measurement of $\sigma_{0}$ can also be converted into a measurement
of the electroweak angle $\sin^{2} \theta_{W}$ according to Eq.\
(\ref{eq:coh-33}), which would provide important complementary insight
into electroweak precision observables at low energies.
In Fig.\ \ref{fig:swplot} we show the sensitivity of this experiment
on $\sin^{2} \theta_{W}$, assuming its SM value at low scale of
$0.238$ (red line). The blue curves represent 3$\sigma$-bounds, solid
for a conservative configuration
$(\sigma_{a},\thinspace\sigma_{f})=(5\%,\thinspace3\%)$
and dashed for a optimistic one
$(\sigma_{a},\thinspace\sigma_{f})=(0.5\%,\thinspace0.1\%)$.
From Fig.\ \ref{fig:swplot} we can see that in the conservative
configuration $\sin^{2} \theta_{W}$ is expected to be measured,
depending on the threshold, to a good precision between 10\% and 20\%,
while in the optimistic configuration, this would be improved
roughly by an order of magnitude. For a threshold of 0.1 keV, the
precision at $3\sigma$ is $\pm 0.0022$, or about 1\%, to be compared with
the dedicated P2 experiment \cite{Berger:2015aaa}, which
aims at a $1\sigma$ precision of 0.13\%.

\subsection{Non-Standard Interactions}

\begin{figure}
\centering

\includegraphics[width=8cm]{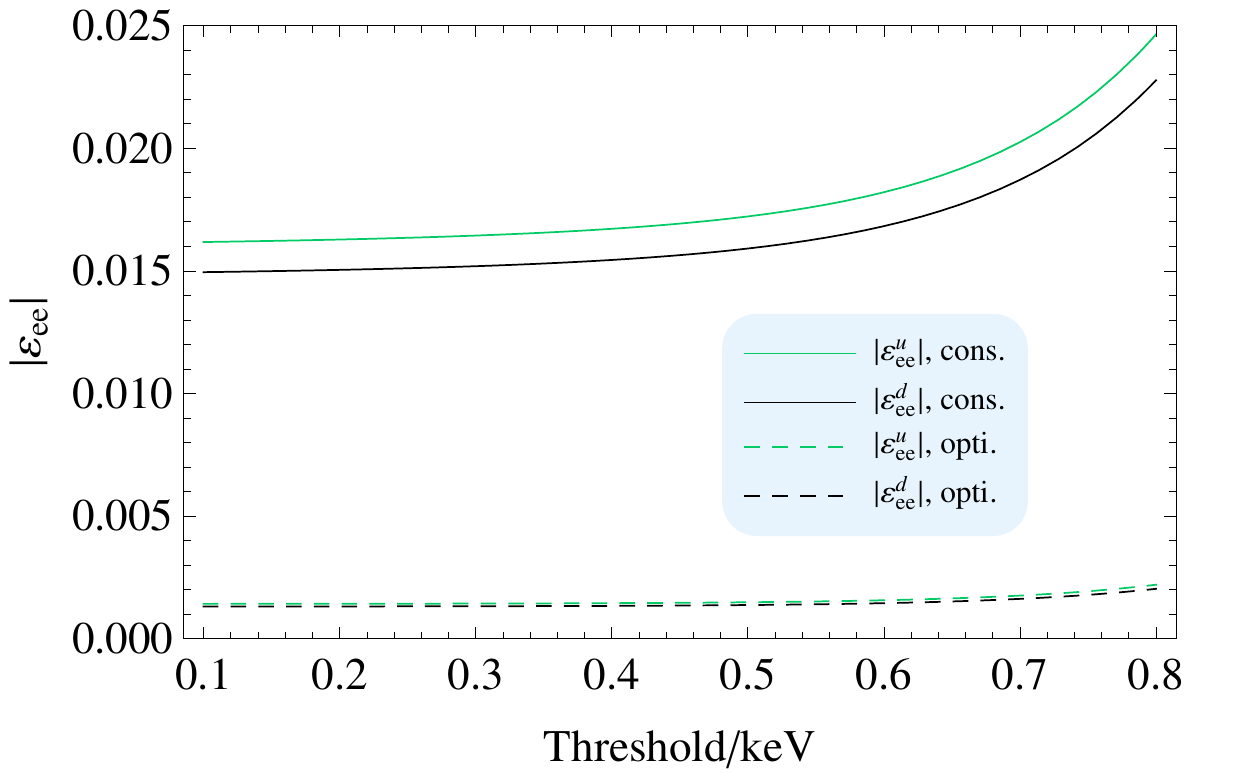}\,\includegraphics[width=8cm]{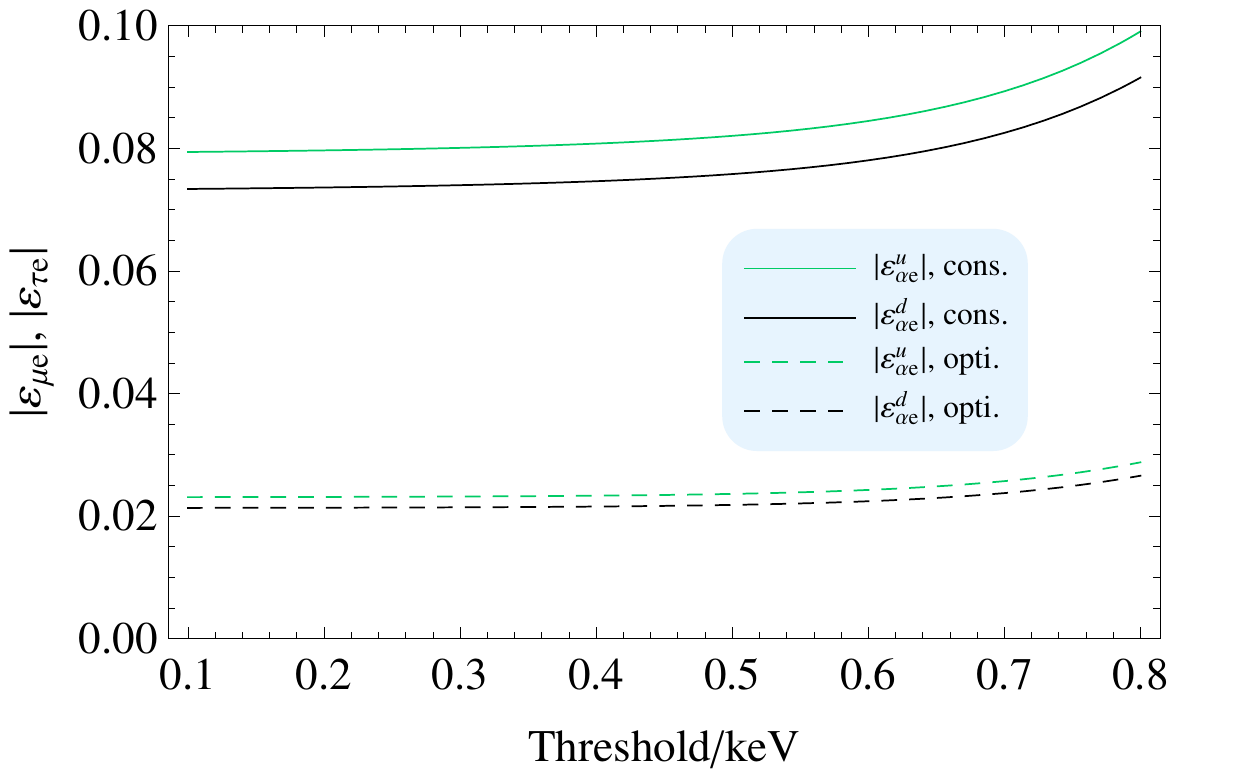}

\caption{\label{fig:NSI-T} Sensitivities on the conventional NSI parameters
(3$\sigma$). The solid and dashed curves are generated for a conservative
and optimistic configuration, respectively.}
\end{figure}

The effect of the conventional NSI, as we have discussed in Sec.\ \ref{sec:NSI},
is merely a correction on the overall factor $\sigma_{0}^{\rm SM}$.
The dependence of $\sigma_{0}^{{\rm {\rm NSI}}}$ on various
$\varepsilon$ parameters has been studied in Sec.\ \ref{sec:NSI} and
was displayed in Fig.\ \ref{fig:NSI-dep}.
We will study here the sensitivities of each $\varepsilon$ individually,
assuming that all others are zero.
The sensitivities on the NSI parameters in both the conservative and
optimistic configurations are presented in Fig.\ \ref{fig:NSI-T}.
The left panel is for $\varepsilon_{ee}^{q}$ with $q=u$ or $d$.
The right panel is for $\varepsilon_{\alpha e}^{q}$ with $\alpha=\mu$
or $\tau$, the cases are indistinguishable.
The left panel of Fig.\ \ref{fig:NSI-T} shows
that the C$\nu N$S  experiment in the conservative
configuration could constrain $\varepsilon_{ee}^{q}$ to order
$10^{-2}$, much better than the current best
bounds given in Eqs.\ (\ref{eq:coh-65}) and (\ref{eq:coh-67}),
which are typically of order $1$. If one
takes the optimistic configuration, then the constraint would reach
the order of $10^{-3}$. The constraints on $\varepsilon_{\mu e}^{q}$
and $\varepsilon_{\tau e}^{q}$, however, are relatively weaker, about
0.07 to $0.10$ (0.02 to 0.03) for the conservative (optimistic) configuration.
This can be easily understood from the form of $Q_{\rm NSI}$ in
Eq.\ (\ref{eq:coh-63}).
For the $\tau$-channel, this is still a significant improvement compared
to the current bound in Eqs.\ (\ref{eq:coh-68}) and (\ref{eq:coh-69})
while for the $\mu$-channel, the current known bound is already very
strong {[}see Eq.\ (\ref{eq:coh-66}){]}; therefore,
even if we take the optimistic
estimation, the constraint would not exceed the known bound.

\begin{figure}
\centering

\includegraphics[width=10cm]{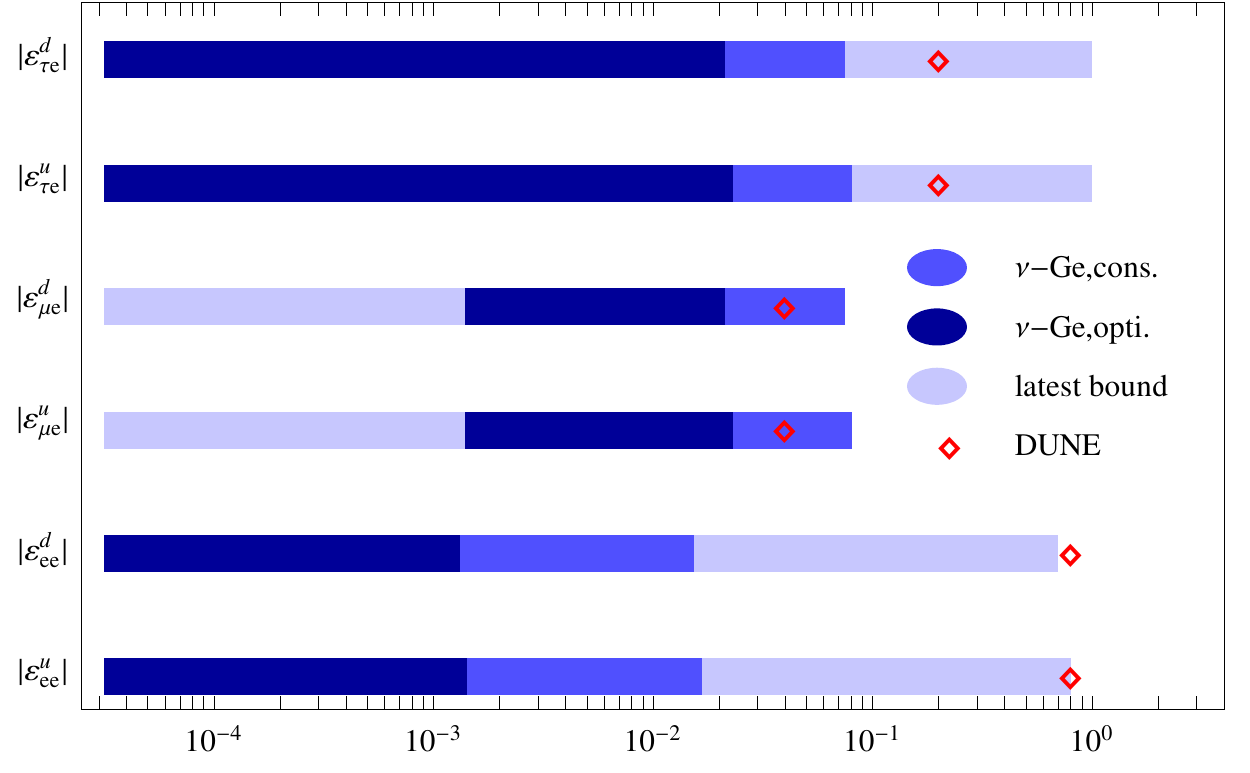}

\caption{\label{fig:NSI-comp} NSI sensitivities compared with the latest known
bounds and the expected constraints from DUNE \cite{deGouvea:2015ndi}.
The label ``latest bound'' indicates the known constraints from
Ref.\ \cite{Davidson:2003ha}. ``$\nu$-Ge, opti.'' and ``$\nu$-Ge,
cons.'' are estimated sensitivities of our assumed 100 kg Ge detector running
for 5 years with optimistic and conservative configurations,
respectively.}
\end{figure}

To summarize  the comparison discussed above, we plot those bounds
in Fig.\ \ref{fig:NSI-comp}. The blue and dark blue bars shows the
3$\sigma$ bounds from our assumed C$\nu N$S experiment
with conservative and optimistic configuration, respectively.
The light blue bars represent the best known bounds from the review
\cite{Davidson:2003ha}, see Eqs.\ (\ref{eq:coh-65}) and (\ref{eq:coh-66}).
We also add the expected bounds \cite{deGouvea:2015ndi}
from the future long-baseline neutrino
experiment DUNE  in the plot. The sensitivity
of DUNE on NSI is based on the modified matter effect of neutrino
oscillations caused by NSI parameters.
The parameter set constrained by DUNE is actually
\begin{equation}
\varepsilon_{\alpha\beta}\equiv\sum_{f=u,d,e}\varepsilon_{\alpha\beta}^{q}\frac{n_{f}}{n_{e}}\approx3\varepsilon_{\alpha\beta}^{u}+3\varepsilon_{\alpha\beta}^{d}+\varepsilon_{\alpha\beta}^{e}\,,\label{eq:coh-77}
\end{equation}
where $n_{f}$ is the number density of the corresponding fermion
$f$. Their relative density ratio $(n_{u}:n_{d}:n_{e})$ is
approximately $(3:3:1)$ in the Earth crust. Focusing on one parameter
at a time, the limits on $\varepsilon_{\alpha\beta}$ from Ref.\
\cite{deGouvea:2015ndi} can be translated into limits on
$\varepsilon_{\alpha\beta}^{q}$. This serves to compare the sensitivities and
is displayed in Fig.\ \ref{fig:NSI-comp}. Even the conservative
configuration improves the bounds on $\varepsilon_{\tau e}^{u,d}$ and
$\varepsilon_{e e}^{u,d}$ considerably beyond current limits and
future DUNE sensitivities.  The limits obtainable in our
benchmark experiment are summarized in Table \ref{tab:nsi}.

\begin{table}
\begin{tabular}{ccccccc}
\hline
 & $\varepsilon_{ee}^{u}$ & $\varepsilon_{ee}^{d}$ & $\varepsilon_{\mu e}^{u}$  & $\varepsilon_{\mu e}^{d}$  & $\varepsilon_{\tau e}^{u}$ & $\varepsilon_{\tau e}^{d}$\tabularnewline
\hline
\hline
Conservative & $1.7\times10^{-2}$ & $1.5\times10^{-2}$ & $8.1\times10^{-2}$ & $7.5\times10^{-2}$ & $8.1\times10^{-2}$ & $7.5\times10^{-2}$\tabularnewline
\hline
Intermediate & $6.0\times10^{-3}$ & $5.5\times10^{-3}$ & $4.8\times10^{-2}$ & $4.4\times10^{-2}$ & $4.8\times10^{-2}$ & $4.4\times10^{-2}$\tabularnewline
\hline
Optimistic & $1.4\times10^{-3}$ & $1.3\times10^{-3}$ & $2.3\times10^{-2}$ & $2.1\times10^{-2}$ & $2.3\times10^{-2}$ & $2.1\times10^{-2}$\tabularnewline
\hline
Latest bound \cite{Davidson:2003ha} & $0.8$ & $0.7$ & $1.4\times10^{-3}$ & $1.4\times10^{-3}$ & $1.0$ & $1.0$\tabularnewline
\hline
DUNE \cite{deGouvea:2015ndi} & $0.8$ & $0.8$ & $0.04$ & $0.04$ & $0.2$ & $0.2$\tabularnewline
\hline
\end{tabular}

\caption{3$\sigma$-bounds on NSI parameters.\label{tab:nsi}}
\end{table}

\subsection{Exotic Neutral Currents}

\begin{figure}
\centering

\includegraphics[width=6cm]{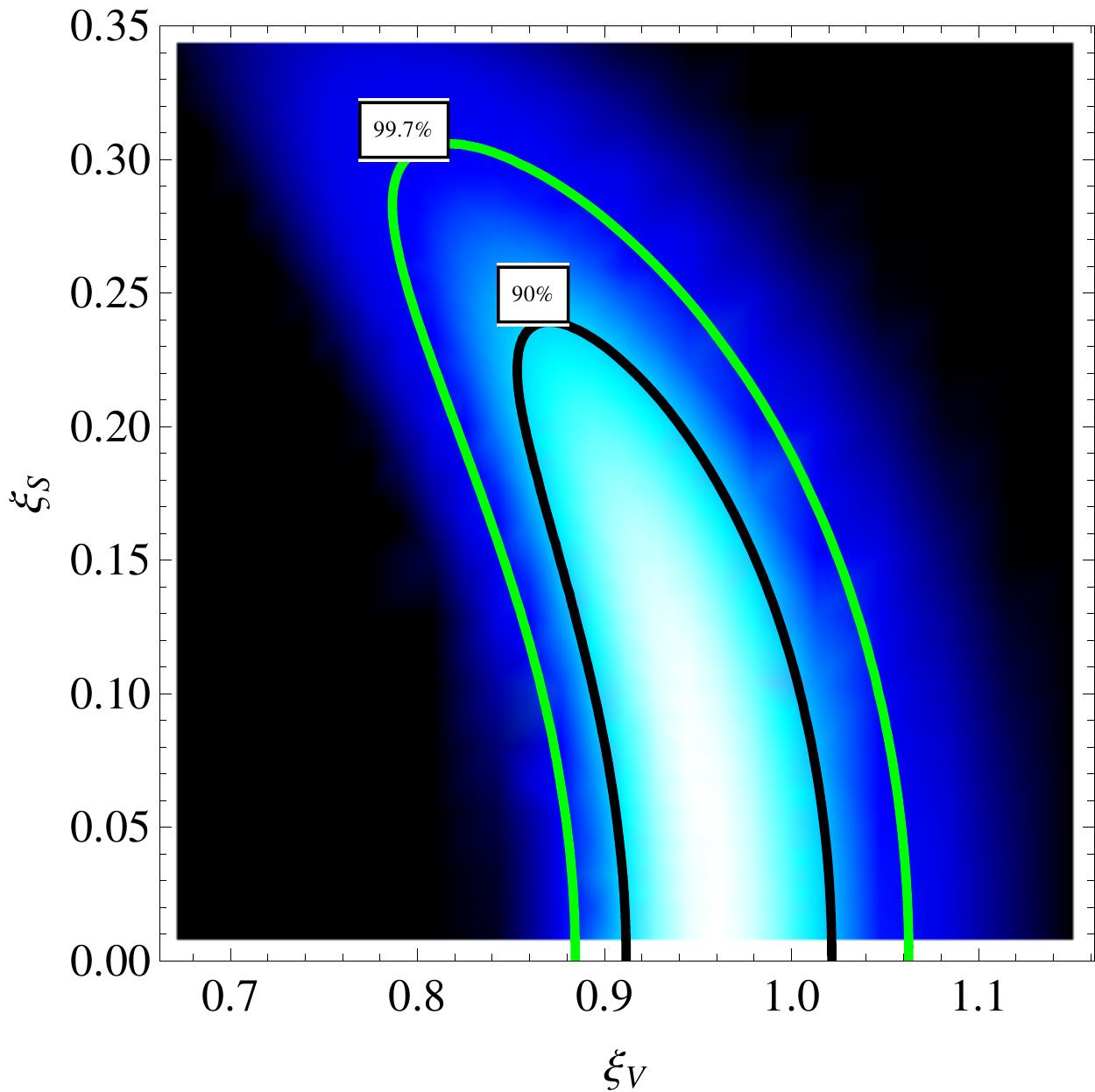}\,\,\includegraphics[width=6cm]{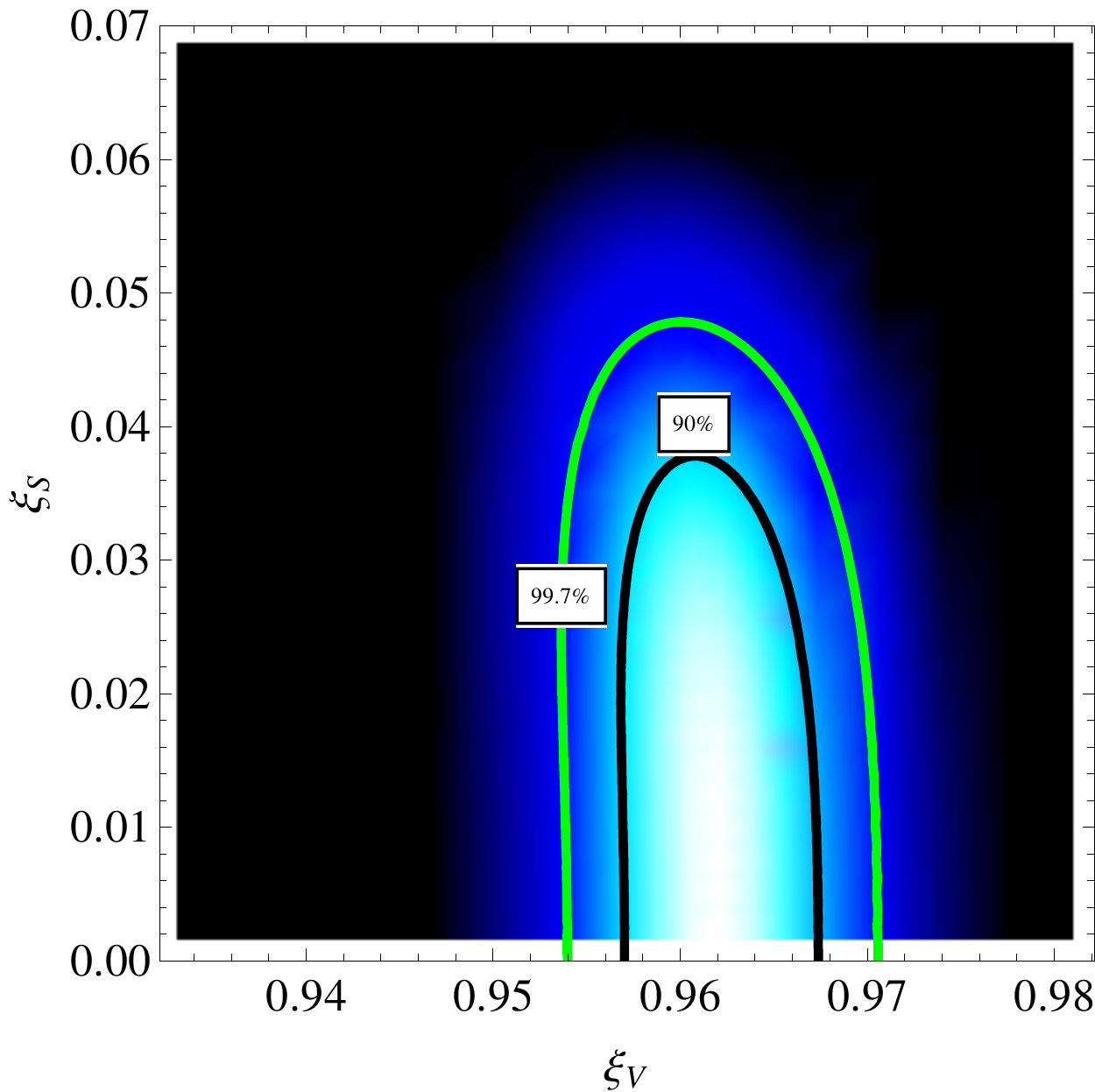}

\includegraphics[width=6cm]{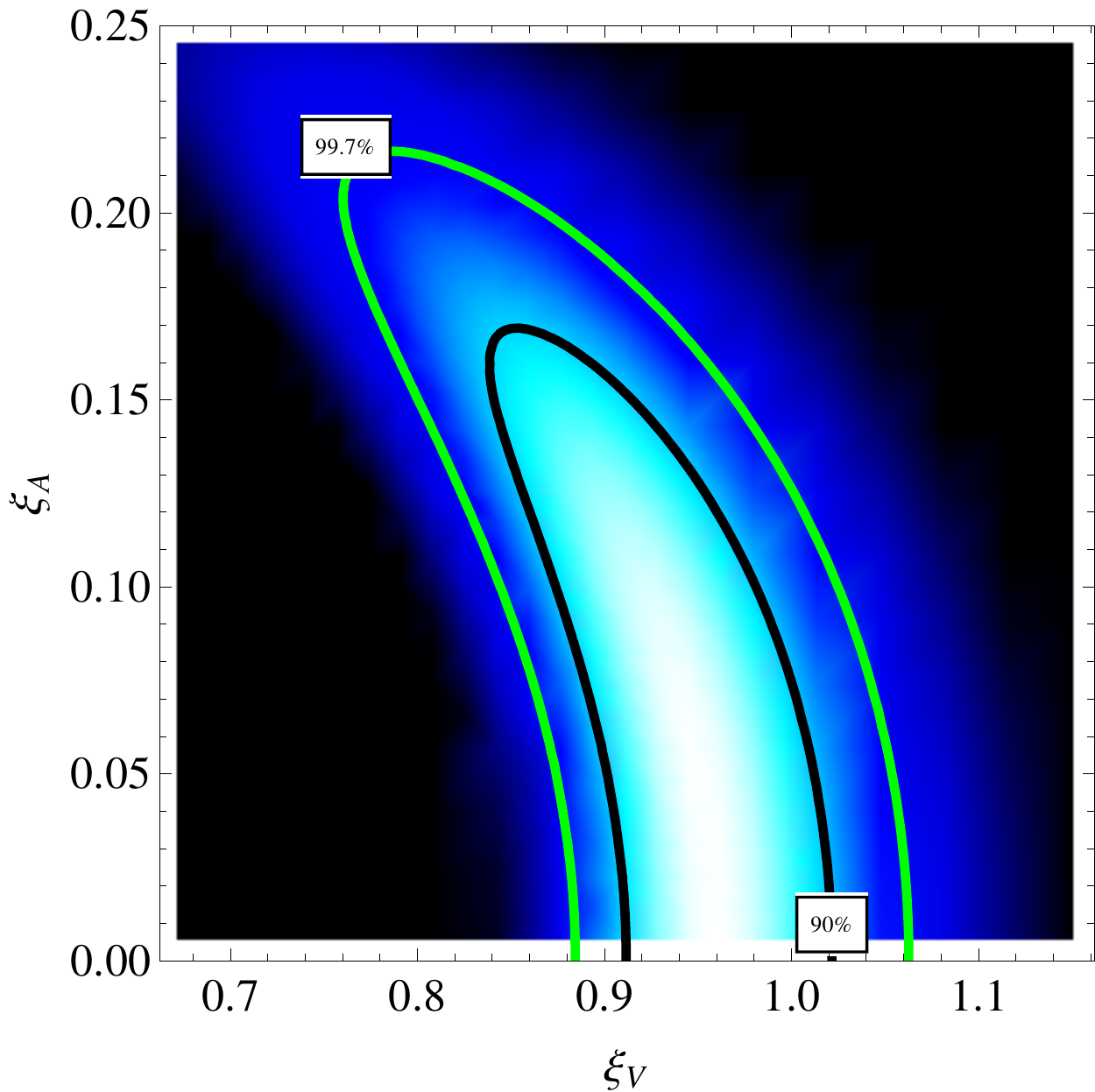}\,\,\includegraphics[width=6cm]{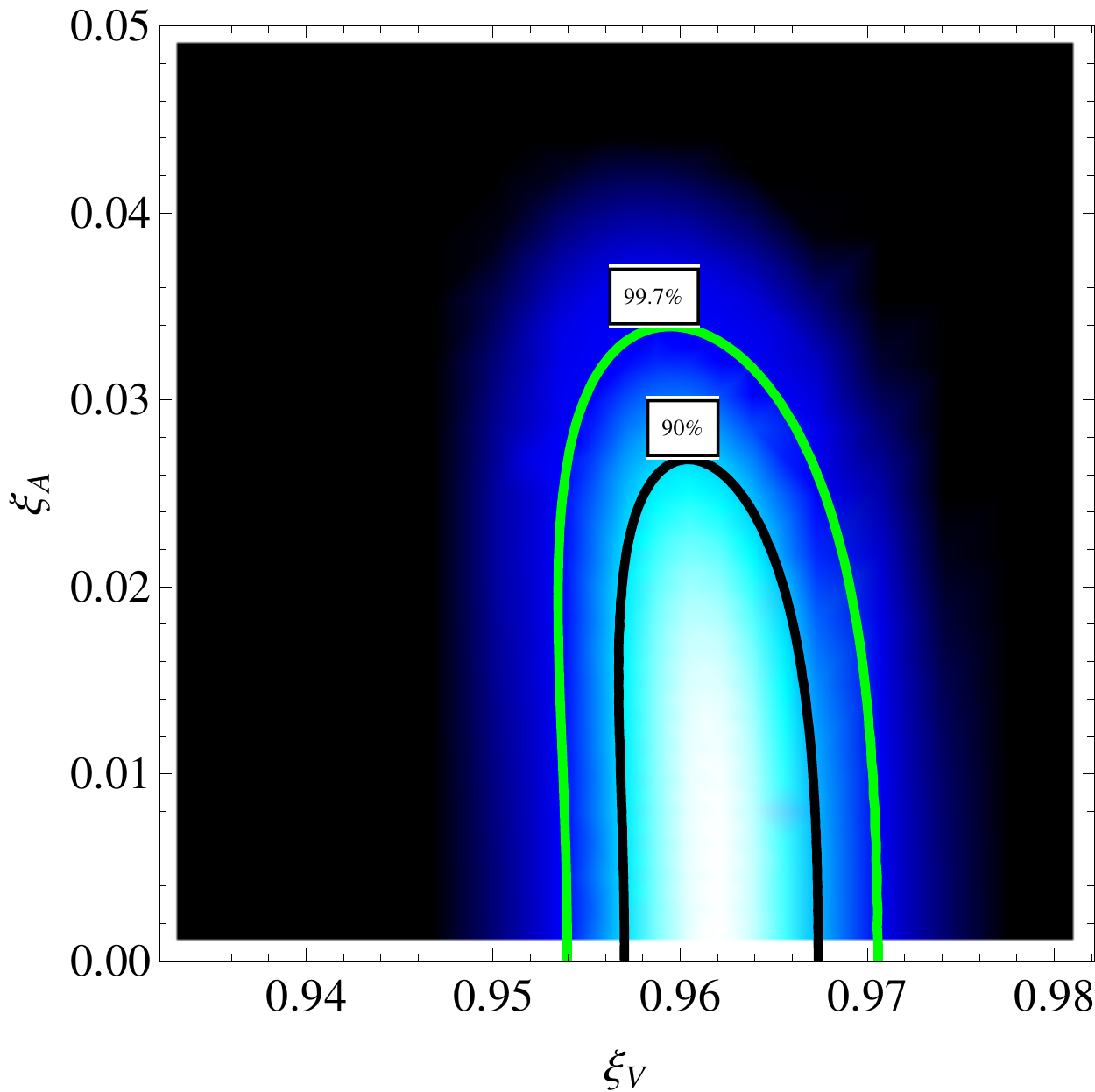}

\includegraphics[width=6cm]{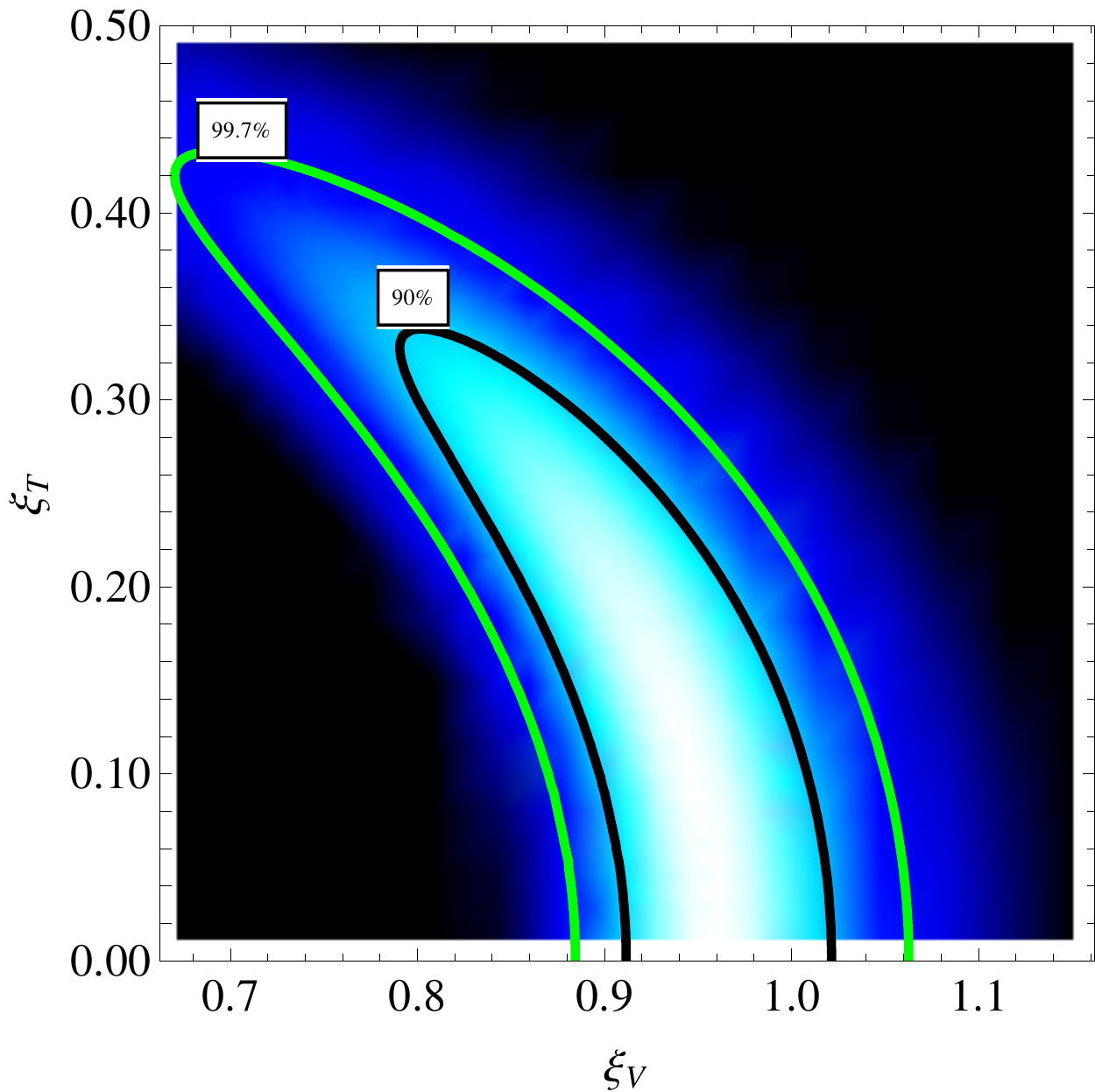}\,\,\includegraphics[width=6cm]{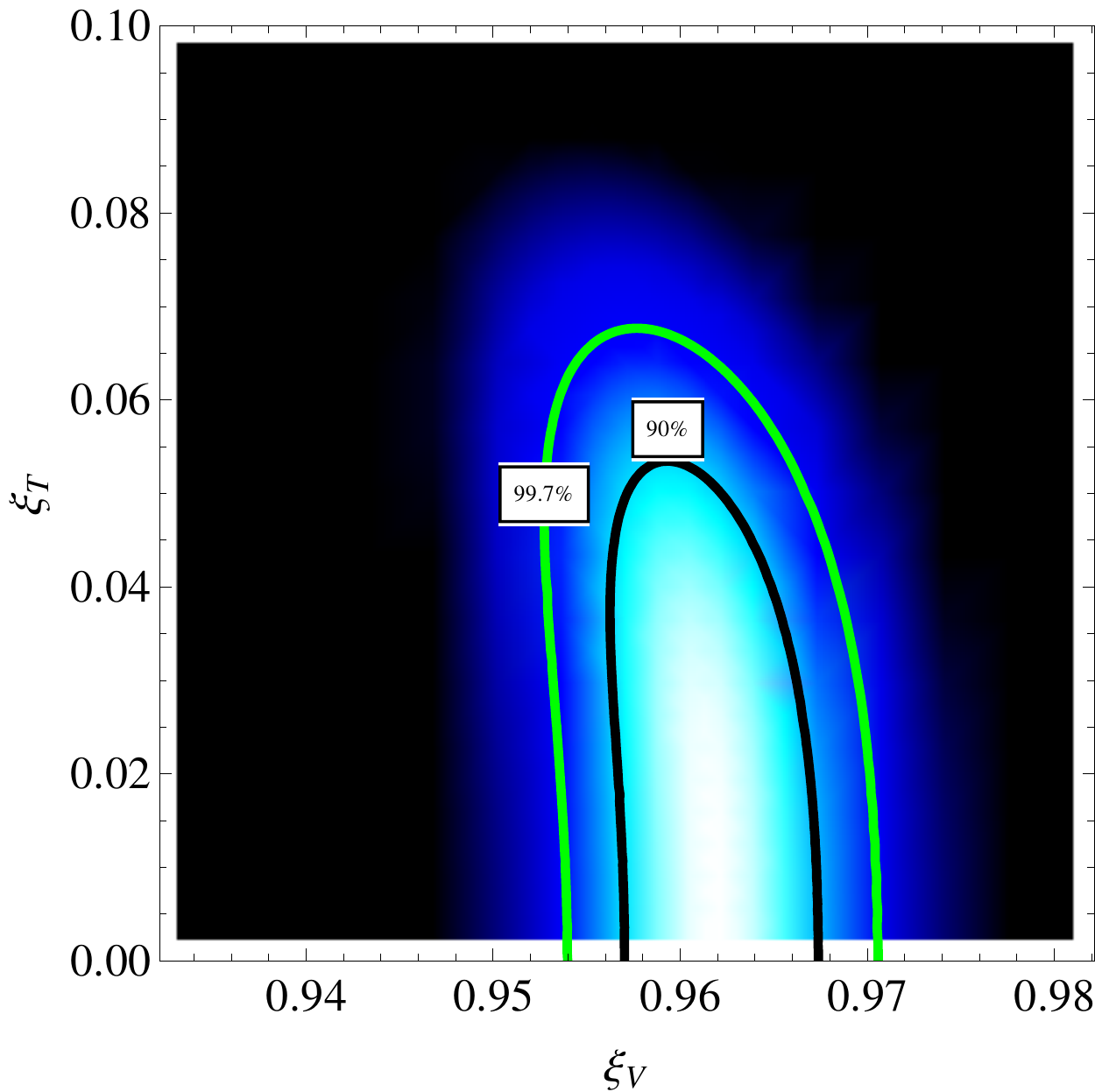}

\caption{\label{fig:exotic-fit}Sensitivity on the exotic neutral currents.
The green and black contours correspond to 99.7\% and 90\% exclusion
bounds. Left (right) panels are generated under the conservative (optimistic)
configuration.}
\end{figure}

Next we shall study the sensitivity on the exotic neutral currents
discussed in Sec.\ \ref{sec:Exotic}. The cross section (\ref{eq:coh-55})
only depends on 5 effective parameters $(\xi_{S},\thinspace\xi_{V},\thinspace\xi_{A},\thinspace\xi_{T},\thinspace R)$, and we
will perform a $\chi^{2}$-fit on those.
Similar to the NSI analysis, we will focus on one type of exotic interactions at a time. However, in our parametrization $\xi_{V}$ is necessarily non-zero as it includes the SM contribution.
So each time we take two non-zero parameters in the fit. One is $\xi_{V}$ and the other one is from exotic couplings.
We also take $R$ to zero since non-zero $R$ would stem from
the interference of scalar and tensor interactions, i.e., would require
the coexistence of two new interactions. Therefore we only
consider three cases, $(\xi_{S},\thinspace\xi_{V})$,
$(\xi_{A},\thinspace\xi_{V})$, and $(\xi_{T},\thinspace\xi_{V})$.

The result is given in Fig.\ \ref{fig:exotic-fit} with both the
conservative (left panels) and optimistic (right panels) configurations
taken. In the conservative configuration,
the sensitivity on $\xi_{V}$ is correlated with the other parameters.
For example, if $\xi_{T}=0$ then $\xi_{V}$ would be only allowed
to stay in the regime $0.88<\xi_{V}<1.06$ at 99.7\% confidence level; if
there is a sizable contribution from the tensor interaction with, say,
$\xi_{T}=0.42$ then $\xi_{V}$ is
allowed to significantly deviate from the SM value, going down to
$0.68$. The correlation could be avoided if the systematic
uncertainties and the threshold are improved to the optimistic configuration,
as is shown in the right panels of Fig.\ \ref{fig:exotic-fit}. The
qualitative explanation is that for large systematic uncertainties,
the sensitivity will mainly depend on the total event number while
the constraint from the spectrum information is not significant. In
this case the tensor interaction will mimic the vector interaction
in the signal, since they both contribute to the total event number.
If the systematic uncertainties are small enough so that the spectrum
is also measured to good accuracy, then the spectrum information could
distinguish the contribution of the tensor interaction from the vector
interaction.  The same argument also applies for the
other two cases $(\xi_{S},\thinspace\xi_{V})$ and $(\xi_{A},\thinspace\xi_{V})$.
Therefore in future C$\nu N$S experiments
reducing the systematic uncertainties is very important to distinguish
signals from new exotic interactions and the SM interaction.
The limits obtainable in our
benchmark experiment are summarized in Table \ref{tab:exo}.

\begin{table}
\begin{tabular}{ccccc}
\hline
 & $\xi_{S}$ & $\xi_{V}$ & $\xi_{A}$ & $\xi_{T}$\tabularnewline
\hline
\hline
Conservative & $0.21$ & $(0.893,\thinspace1.048)$ & $0.14$ & $0.25$\tabularnewline
\hline
Intermediate & $0.11$ & $(0.934,\thinspace0.993)$ & $7.8\times10^{-2}$ & $0.14$\tabularnewline
\hline
Optimistic & $4.4\times10^{-2}$ & $(0.955,\thinspace0.970)$ & $3.1\times10^{-2}$ & $5.9\times10^{-2}$\tabularnewline
\hline
\end{tabular}

\caption{3$\sigma$-bounds on exotic neutral current parameters,
see Eq.\ (\ref{eq:coh-55}). The SM value of $\xi_{V}$ is 0.962.
\label{tab:exo}}
\end{table}

\section{\label{sec:Conclusion}Conclusion}
Future coherent neutrino-nucleus scattering experiments will provide
exciting new data to test Standard Model and new neutrino physics
to unprecedented accuracy. We have assumed here a future experiment with
a low threshold (down to 0.1 keV nuclear recoil) Germanium detector,
with experimental benchmark numbers of
500 kg $\times$ years $\times$ GW reactor neutrinos and a baseline of 10 m.
We firmly believe that such a setup is achievable within the next decade,
and it will provide event numbers of the order of $10^5$.
Constraints on neutrino non-standard interactions and exotic
neutral current interactions were evaluated.  The expected sensitivities were shown to reach
percent and permille level when compared to Fermi interaction, significantly better than
expected constraints from oscillation experiments.
We have demonstrated that such
comparably compact coherent scattering experiments open a new window into exciting physics
and should be pursued actively.

\begin{acknowledgments}
We thank Carlos Yaguna and Thomas Rink for many useful discussions.  WR  is supported
by the DFG with grant RO 2516/6-1 in the Heisenberg Programme.
\end{acknowledgments}

\appendix

\section{\label{sec:Cross-section}Cross Section Calculation of coherent
$\nu-N$ Scattering in the Standard Model}

In the SM, the neutral current (NC) is
\begin{eqnarray}
J_{{\rm NC}}^{\mu} & = & 2\sum_{f}g_{L}^{f}\overline{f_{L}}\gamma^{\mu}f_{L}+g_{R}^{f}\overline{f_{R}}\gamma^{\mu}f_{R}\label{eq:coh-1}\\
 & = & \sum_{f}\overline{f}\gamma^{\mu}(g_{V}^{f}-g_{A}^{f}\gamma^{5})f\,,\label{eq:coh-2}
\end{eqnarray}
where $f$ stands for all elementary fermions in the SM and $f_{L,R}$
are their left/right-handed components,
\begin{equation}
f_{L}=\frac{1-\gamma^{5}}{2}f\,,~ f_{R}=\frac{1+\gamma^{5}}{2}f\,.\label{eq:coh-3}
\end{equation}
Here $g_{L,R}^{f}$ are determined by the quantum numbers of the corresponding
fermions under $SU(2)_{L}\times U(1)_{Y}$:
\begin{eqnarray}
 &  & g_{L}^{\nu}=\frac{1}{2}\,,~ g_{R}^{\nu}=0,\thinspace g_{L}^{e}=-\frac{1}{2}+s_{W}^{2}\,,~ g_{R}^{e}=s_{W}^{2}\,.\label{eq:coh-4}\\
 &  & g_{L}^{u}=\frac{1}{2}-\frac{2}{3}s_{W}^{2}\,,~ g_{R}^{u}=-\frac{2}{3}s_{W}^{2}\,,~ g_{L}^{d}=-\frac{1}{2}+\frac{1}{3}s_{W}^{2}\,,~ g_{R}^{d}=\frac{1}{3}s_{W}^{2}\,.\label{eq:coh-5}
\end{eqnarray}
The vector/axial couplings $g_{V,A}^{f}$ are defined as
\begin{equation}
g_{V}^{f}=g_{L}^{f}+g_{R}^{f}\,,~g_{A}^{f}=g_{L}^{f}-g_{R}^{f}\,.\label{eq:coh-6}
\end{equation}
At low energies the effective NC interaction is
\begin{equation}
{\cal L}_{{\rm NC}}=\frac{G_{F}}{\sqrt{2}}J_{\rm NC}^{\mu}J_{{\rm NC}\mu}\,, \label{eq:coh-7}
\end{equation}
therefore the amplitude of the coherent $\nu-N$ scattering is
\begin{equation}
i{\cal M}(\nu+N\rightarrow\nu+N)=-i\sqrt{2}G_{F}\langle N(k_{2})|J_{\rm NC}^{\mu}|N(p_{2})\rangle\langle\nu(k_{1})|J_{\rm NC\mu}|\nu(p_{1})\rangle\,,\label{eq:coh-8}
\end{equation}
where $p_{1}$, $k_{1}$, $p_{2}$, $k_{2}$ are the momenta of the initial
neutrino, final neutrino, initial nucleus and final nucleus, respectively.

The matrix element $\langle N(k_{2})|J_{{\rm NC}}^{\mu}|N(p_{2})\rangle$
only depends on the quark sector in $J_{{\rm NC}}^{\mu}$
since the nucleus is a bound state of many $u$ and $d$ quarks. So
we have
\begin{eqnarray}
\langle N|J_{{\rm NC}}^{\mu}|N\rangle & = & g_{L}^{u}\langle N|\overline{u_{L}}\gamma^{\mu}u_{L}|N\rangle+g_{R}^{u}\langle N|\overline{u_{R}}\gamma^{\mu}u_{R}|N\rangle\nonumber \\
 & + & g_{L}^{d}\langle N|\overline{d_{L}}\gamma^{\mu}d_{L}|N\rangle+g_{R}^{d}\langle N|\overline{d_{R}}\gamma^{\mu}d_{R}|N\rangle\,.\label{eq:coh-10}
\end{eqnarray}
Assuming that the nucleus does not violate parity, we have
\begin{equation}
\langle N|\overline{u_{L}}\gamma^{\mu}u_{L}|N\rangle=\langle N|\overline{u_{R}}\gamma^{\mu}u_{R}|N\rangle,\thinspace\langle N|\overline{d_{L}}\gamma^{\mu}d_{L}|N\rangle=\langle N|\overline{d_{R}}\gamma^{\mu}d_{R}|N\rangle.\label{eq:coh-11}
\end{equation}
Note that generally $|N\rangle$ does not have to respect parity symmetry.
For example, if the whole nucleus is a spin-$1/2$ fermion
then it is impossible for $|N\rangle$ to be
invariant under the parity transformation which would flip the orientation
of the spin. Even if the nucleus is a spin-0 particle, for the $u$
quarks the number of spin-up could be different from the number of
spin-down\footnote{Besides, the distribution of protons and neutrons in a nucleus is
not spherical, though it tends to be more spherical with increasing atomic
number. }. However, for a nucleus with a large mass number $A$, it contains
many $u$ and $d$ quarks so that statistically we expect that they
form a large object (the nucleus) that approximately respects parity.

Another relation we will use is
\begin{equation}
\frac{\langle N|\overline{u}\gamma^{\mu}u|N\rangle}{\langle N|\overline{d}\gamma^{\mu}d|N\rangle}=\frac{2Z+N}{2N+Z},\label{eq:coh}
\end{equation}
where $N$ is the number of neutrons and $Z$ the number of protons
in the nucleus.
Note that in a nucleus with $N$ neutrons and $Z$
protons, the numbers of $u$ and $d$ quarks are $2Z+N$ and $2N+Z$
respectively. Their ratio must be identical to the ratio of the above
matrix elements if all the quarks are free particles. Since the strong
interaction   can not distinguish $u$ quarks and
$d$ quarks, we assume that this relation holds for the bound quarks
in the nucleus as well.

From Eq.\ (\ref{eq:coh}) we can write down
\begin{equation}
\langle N|\overline{u}\gamma^{\mu}u|N\rangle=(2Z+N)f^{\mu},\thinspace\langle N|\overline{d}\gamma^{\mu}d|N\rangle=(2N+Z)f^{\mu}\,,\label{eq:coh-12}
\end{equation}
where $f^{\mu}$ can be determined by the electromagnetic property
of the nucleus. Let us first consider the electromagnetic current
\begin{equation}
J_{{\rm EM}}^{\mu}=\frac{2}{3}\overline{u}\gamma^{\mu}u+\frac{-1}{3}\overline{d}\gamma^{\mu}d\,.\label{eq:coh-13}
\end{equation}
From the Feynman rules of a complex scalar field with a gauged $U(1)$
symmetry we know that the interaction vertex of the gauge boson with
the scalar field should be proportional to $(p_{2}+k_{2})^{\mu}$.
Therefore we have
\begin{equation}
\langle N(k_{2})|J_{\rm EM}^{\mu}|N(p_{2})\rangle=(p_{2}+k_{2})^{\mu}Q_{{\rm nucl}}F(q^{2})\,,\label{eq:coh-16}
\end{equation}
where $Q_{{\rm nucl}}=Z$ is the electric charge of the nucleus and
$q^{\mu}$ is the momentum transfer, defined as
$q^{\mu}\equiv k_{2}^{\mu}-p_{2}^{\mu}$. 
From Eqs.\ (\ref{eq:coh-12}), (\ref{eq:coh-13}) and (\ref{eq:coh-16}),
we obtain the form of $f^{\mu}$:
\begin{equation}
f^{\mu}=(p_{2}+k_{2})^{\mu}F(q^{2})\,.\label{eq:coh-17}
\end{equation}
For very soft photons ($q^{2}\rightarrow0$) in the electromagnetic
interaction, the nucleus radius $r_{{\rm nucl}}$ is much smaller
than the electromagnetic wavelength so that it can be treated as a
point-like particle, with electric charge $Q_{{\rm nucl}}$. Therefore
for a very soft momentum transfer we have
\begin{equation}
F(q^{2}\ll 1/r_{{\rm nucl}}^{2})\approx1\,.\label{eq:coh-15}
\end{equation}
With Eqs.\ (\ref{eq:coh-11}), (\ref{eq:coh-12}) and (\ref{eq:coh-17}),
Eq.\ (\ref{eq:coh-10}) can now be written as
\begin{eqnarray}
\langle N(k_{2})|J_{\rm NC}^{\mu}|N(p_{2})\rangle & = & F(q^{2})(p_{2}+k_{2})^{\mu}\left[(2Z+N)g_{V}^{u}+(2N+Z)g_{V}^{d}\right]\nonumber \\
 & = & F(q^{2})(p_{2}+k_{2})^{\mu}\left[Zg_{V}^{p}+Ng_{V}^{n}\right],\label{eq:coh-9}
\end{eqnarray}
where
\begin{equation}
g_{V}^{p}=\frac{1}{2}-2s_{W}^{2}\,~ g_{V}^{n}=-\frac{1}{2}\,.\label{eq:coh-18}
\end{equation}
Some references \cite{Scholberg:2005qs,Anderson:2011bi} define the
weak charge $Q_{W}$ which is
\begin{equation}
Q_{W}=-2(Zg_{V}^{p}+Ng_{V}^{n})=N-(1-4s_{W}^{2})Z\,.\label{eq:coh-19}
\end{equation}
Now we can continue the evaluation of Eq.\ (\ref{eq:coh-8})
\begin{equation}
i{\cal M}^{ss'}(\nu+N\rightarrow\nu+N)=i\frac{\sqrt{2}}{2}G_{F}Q_{W}F(q^{2})g_{L}^{\nu}(p_{2}+k_{2})^{\mu}\overline{v}^{s}(p_{1})\gamma^{\mu}(1-\gamma^{5})v^{s'}(k_{1})\,,\label{eq:coh-20}
\end{equation}
where $s$ and $s'$ are the helicities of the initial neutrino and
final neutrino, both left-handed. When computing $|i{\cal M}|^{2}$
we can also use the trace technology since the right-handed case should
vanish due to the $V-A$ coupling of neutrinos in Eq.\ (\ref{eq:coh-20}),
\begin{eqnarray}
|i{\cal M}|^{2} & = & \sum_{ss'}|i{\cal M}^{ss'}|^{2}\,.\label{eq:coh-22}
\end{eqnarray}
One can evaluate it immediately\footnote{Some kinetic relations are needed in the calculation, including $q^{2}=2MT$
and $p_{1}\cdot q=-p_{2}\cdot q=q^{2}/2$. The former is from $p_{2}\cdot k_{2}=M(M+T)$
and the latter is from the on-shell conditions of $k_{1}$ and $k_{2}$.}:
\begin{equation}
|i{\cal M}|^{2}=32G_{F}^{2}Q_{W}^{2}F^{2}(g_{L}^{\nu})^{2}M^{2}E_{\nu}^{2}\left(1-\frac{T}{E_{\nu}}-\frac{MT}{2E_{\nu}^{2}}\right),\label{eq:coh-21}
\end{equation}
where $M$ is the nucleus mass and $E_{\nu}$ the neutrino energy;
$T$ is the recoil energy of the nucleus, which can be related to
$c_{\theta}\equiv\cos\theta$, where $\theta$ is defined as the scattering
angle between the momenta of the initial neutrino and final nucleus,
\begin{equation}
T=\frac{2ME_{\nu}^{2}c_{\theta}^{2}}{(M+E_{\nu})^{2}-E_{\nu}^{2}c_{\theta}^{2}}\,.\label{eq:coh-23}
\end{equation}
For a given value of $E_{\nu}$, the maximal recoil energy $T_{{\rm max}}$
is reached at $\theta=0$:
\begin{equation}
T_{{\rm max}}(E_{\nu})=\frac{2E_{\nu}^{2}}{M+2E_{\nu}}\,.\label{eq:coh-29}
\end{equation}
In the form factor $F(q^{2})$, $q^{2}$ is needed, which can be expressed
in terms of $T$ as $ q^{2}=-2MT$.

The differential cross section in the laboratory frame  is
\begin{equation}
\frac{d\sigma}{dc_{\theta}}=\frac{|{\cal M}|^{2}}{8\pi}\frac{c_{\theta}(E_{\nu}+M)^{2}}{\left[(M+E_{\nu})^{2}-E_{\nu}^{2}c_{\theta}^{2}\right]^{2}}\,,\label{eq:coh-24}
\end{equation}
or
\begin{equation}
\frac{d\sigma}{dT}=\frac{|{\cal M}|^{2}}{32\pi ME_{\nu}^{2}}.\label{eq:coh-25}
\end{equation}
With the result in Eq.\ (\ref{eq:coh-21}) we have
\begin{equation}
\frac{d\sigma}{dT}=\frac{G_{F}^{2}(2g_{L}^{\nu}Q_{W})^{2}F^{2}(q^{2})}{4\pi}M\left(1-\frac{T}{E_{\nu}}-\frac{MT}{2E_{\nu}^{2}}\right).\label{eq:coh-27}
\end{equation}
We have finally arrived at the expression of the SM cross section in
Eqs.\ (\ref{eq:coh-34})-(\ref{eq:coh-29-1}).

Finally, considering that $E_{\nu}\ll M$, many expressions can be simplified
under this approximation. From Eqs.\ (\ref{eq:coh-23})
and (\ref{eq:coh-29}) we have
\begin{equation}
T\approx T_{{\rm max}}c_{\theta}^{2}\label{eq:coh-30}
\end{equation}
and
\begin{equation}
1-\frac{T}{E_{\nu}}-\frac{MT}{2E_{\nu}^{2}}\approx\sin^{2}\theta+{\cal O}\left(\frac{E_{\nu}^{2}}{M^{2}}\right),\label{eq:coh-28}
\end{equation}
which gives
\begin{eqnarray}
\frac{d\sigma}{dT} & \approx & \frac{\sigma_{0}^{{\rm SM}}}{M}\sin^{2}\theta\,.\label{eq:coh-32}
\end{eqnarray}

\section{\label{sec:What-if}What if $N$ is a Spin-1/2 or Spin-1 Particle?}
We may ask whether non-zero spins have a significant effect on
the calculation presented above or not.
An intuitive estimation is that it should
be only a weak effect. The reason is that a large nucleus contains
many spin-1/2 fermions, i.e.\ protons and neutrons. They
form the nucleus in which the proton and neutron spins almost cancel.
If one proton flips its spin, the nucleus spin would
be changed e.g.\ from $0$ to 1. Since we expect that
the coherent $\nu-N$ scattering is insensitive to the status of a
single proton inside the nucleus, we suspect that there
should be no significant difference between zero and non-zero spins,
as long as the non-zero spin is not very high.

For a spin-1/2 nucleus, Eq.\ (\ref{eq:coh-9}) is modified
to
\begin{equation}
\langle N(k_{2},\thinspace r')|J_{{\rm NC}}^{\mu}|N(p_{2},\thinspace r)\rangle=F(q^{2})\overline{u}^{r'}(k_{2})\gamma^{\mu}u^{r}(p_{2})\left[Zg_{V}^{p}+Ng_{V}^{n}\right],\label{eq:coh-38}
\end{equation}
where $\overline{u}^{r'}(k_{2})$ and $\overline{u}^{r}(p_{2})$ denote
the finial and initial states of the Dirac particle, i.e., the spin-1/2
nucleus. Then Eq.\ (\ref{eq:coh-20}) is changed to
\begin{equation}
i{\cal M}^{r'rss'}(\nu+N\rightarrow\nu+N)=i\frac{\sqrt{2}}{2}G_{F}Q_{W}F(q^{2})g_{L}^{\nu}\left[\overline{u}^{r'}(k_{2})\gamma^{\mu}u^{r}(p_{2})\right]\left[\overline{u}^{s'}(k_{1})\gamma^{\mu}(1-\gamma^{5})u^{s}(p_{1})\right].\label{eq:coh-39}
\end{equation}
The above amplitude is for neutrinos while for antineutrinos it should
be
\begin{equation}
i{\cal M}^{r'rss'}(\overline{\nu}+N\rightarrow\overline{\nu}+N)=i\frac{\sqrt{2}}{2}G_{F}Q_{W}F(q^{2})g_{L}^{\nu}\left[\overline{u}^{r'}(k_{2})\gamma^{\mu}u^{r}(p_{2})\right]\left[\overline{v}^{s}(p_{1})\gamma^{\mu}(1-\gamma^{5})v^{s'}(k_{1})\right].\label{eq:coh-39-1}
\end{equation}
Eq.\ (\ref{eq:coh-39}) and Eq.\ (\ref{eq:coh-39-1}) essentially
give the same $|{\cal M}|^{2}$ and thus the same cross section, as
one can check by direct computation. The reason is due to the assumption
that the interaction of the nucleus with the $Z$ boson is parity-conserved.
If there is axial current in Eq.\ (\ref{eq:coh-38}), i.e., a $\gamma^{\mu}\gamma^{5}$
between $\overline{u}^{r'}(k_{2})$ and $u^{r}(p_{2})$ then the $\nu N$
and $\overline{\nu}N$ cross sections would be different.

After evaluating the traces of the Dirac matrices in the amplitude,
we get
\begin{equation}
|{\cal M}|^{2}=\sum_{ss'}\frac{1}{2}\sum_{rr'}|i{\cal M}^{r'rss'}|^{2}=32G_{F}^{2}Q_{W}^{2}F^{2}(g_{L}^{\nu})^{2}M^{2}E_{\nu}^{2}\left(1-\frac{T}{E_{\nu}}-\frac{MT}{2E_{\nu}^{2}}+\frac{T^{2}}{2E_{\nu}^{2}}\right),\label{eq:coh-40}
\end{equation}
and then
\begin{equation}
\left.\frac{d\sigma}{dT}\right|_{\mbox{spin-1/2}}=\frac{G_{F}^{2}(2g_{L}^{\nu}Q_{W})^{2}F^{2}(q^{2})}{4\pi}M\left(1-\frac{T}{E_{\nu}}-\frac{MT}{2E_{\nu}^{2}}+\frac{T^{2}}{2E_{\nu}^{2}}\right).\label{eq:coh-70}
\end{equation}
This is the result for a spin-0 nucleus plus small negligible corrections, see
Eq.\ (\ref{eq:coh-41}).

\section{\label{sec:formfactor}Relations of $(C_{a},\thinspace\overline{D}_{a})$
with $(C_{a}^{(q)},\thinspace\overline{D}_{a}^{(q)})$}

In Sec.\ \ref{sec:Exotic} when discussing exotic neutral currents
we defined the nucleus couplings $(C_{a},\thinspace\overline{D}_{a})$
and the quark couplings $(C_{a}^{(q)},\thinspace\overline{D}_{a}^{(q)})$.
In this appendix we will derive the relations of $(C_{a},\thinspace\overline{D}_{a})$
to $(C_{a}^{(q)},\thinspace\overline{D}_{a}^{(q)})$ by comparing
the scattering amplitudes.

Starting from the fundamental Lagrangian (\ref{eq:coh-35}), we
can write down the amplitude
\begin{eqnarray}
i{\cal M}^{s'sr'r} & = & -i\frac{G_{F}}{\sqrt{2}}\overline{v}^{s'}(p_{1})P_{R}\Gamma^{a}v^{s}(k_{1})\langle\Gamma^{a}\rangle_{N}^{r'r}\,,\label{eq:coh-79}
\end{eqnarray}
where
\begin{equation}
\langle\Gamma^{a}\rangle_{N}^{r'r}\equiv\langle N(k_{2},r')|\sum_{q=u,d}\overline{q}\Gamma^{a}(C_{a}^{(q)}+\overline{D}_{a}^{(q)}i\gamma^{5})q|N(p_{2},r)\rangle\,.\label{eq:coh-87}
\end{equation}
To compute the amplitude we need to know $\langle N|\overline{q}\Gamma^{a}q|N\rangle$.
Similar to Eq.\ (\ref{eq:coh}), here we also assume that
\begin{equation}
\frac{\langle N|\overline{u}\Gamma^{a}u|N\rangle}{\langle N|\overline{d}\Gamma^{a}d|N\rangle}=\frac{n_{u}}{n_{d}}\,,\label{eq:coh-80}
\end{equation}
which enables us to define
\begin{equation}
F^{a}\equiv\frac{\langle N|\overline{u}\Gamma^{a}u|N\rangle}{n_{u}}=\frac{\langle N|\overline{d}\Gamma^{a}d|N\rangle}{n_{d}}\,.\label{eq:coh-81}
\end{equation}
For instance,
generalizing Eq.\ (\ref{eq:coh-38}), a scalar interaction of down quarks
$\langle N|\overline{d} d|N\rangle$, can result in
a term $\overline{u} u$ or $\overline{u} \gamma_5 u$, where $u$
is a Dirac spinor. Both terms come with a form factor, and we have neglected
terms involving momenta. This implies that
\begin{equation}
F^{S}=f_{SS}\overline{u}^{r'}(k_{2})\Gamma^{S}u^{r}(p_{2})+f_{SP}\overline{u}^{r'}(k_{2})\Gamma^{P}u^{r}(p_{2})\,. \label{eq:coh-82}
\end{equation}
In analogy, we can write the other terms as
\begin{equation}
F^{P}=f_{PS}\overline{u}^{r'}(k_{2})\Gamma^{P}u^{r}(p_{2})+f_{PP}\overline{u}^{r'}(k_{2})\Gamma^{S}u^{r}(p_{2})\,,\label{eq:coh-83}
\end{equation}
\begin{equation}
F^{V}=f_{VV}\overline{u}^{r'}(k_{2})\Gamma^{V}u^{r}(p_{2})+f_{VA}\overline{u}^{r'}(k_{2})\Gamma^{A}u^{r}(p_{2})\,,\label{eq:coh-84}
\end{equation}
\begin{equation}
F^{A}=f_{AV}\overline{u}^{r'}(k_{2})\Gamma^{A}u^{r}(p_{2})+f_{AA}\overline{u}^{r'}(k_{2})\Gamma^{V}u^{r}(p_{2}) \,,\label{eq:coh-85}
\end{equation}
\begin{equation}
F^{T}=f_{T}\overline{u}^{r'}(k_{2})\Gamma^{T}u^{r}(p_{2})+f_{T'}\overline{u}^{r'}(k_{2})\Gamma^{T}(i\gamma^{5})u^{r}(p_{2})\,,\label{eq:coh-86}
\end{equation}
where all the $f$ are form factors. We will not address the calculation
of form factors in this paper, see Refs.\ \cite{Belanger:2008sj,Bishara:2016hek}
and references therein.

From the definition (\ref{eq:coh-81}),
we can express $\langle\Gamma^{a}\rangle_{N}$
in terms of $F^{a}$:
\begin{equation}
\langle\Gamma^{S}\rangle_{N}=\sum_{q=u,d}n_{q}(C_{S}^{(q)}F^{S}+\overline{D}_{S}^{(q)}F^{P})\,,\label{eq:coh-88}
\end{equation}
\begin{equation}
\langle\Gamma^{P}\rangle_{N}=\sum_{q=u,d}n_{q}(C_{P}^{(q)}F^{P}-\overline{D}_{P}^{(q)}F^{S})\,,\label{eq:coh-89}
\end{equation}
\begin{equation}
\langle\Gamma^{V}\rangle_{N}=\sum_{q=u,d}n_{q}(C_{V}^{(q)}F^{V}+i\overline{D}_{V}^{(q)}F^{A})\,,\label{eq:coh-90}
\end{equation}
\begin{equation}
\langle\Gamma^{A}\rangle_{N}=\sum_{q=u,d}n_{q}(C_{A}^{(q)}F^{A}+i\overline{D}_{A}^{(q)}F^{V})\,,\label{eq:coh-91}
\end{equation}
\begin{equation}
\langle\Gamma^{T}\rangle_{N}^{\mu\nu}=\sum_{q=u,d}n_{q}\left[C_{T}^{(q)}(F^{T})^{\mu\nu}-\frac{1}{2}\epsilon^{\mu\nu\rho\sigma}\overline{D}_{T}^{(q)}(F^{T})_{\rho\sigma}\right].\label{eq:coh-92}
\end{equation}
We have suppressed the spin indices $(r,\thinspace r',\ldots)$
and Lorentz indices $(\mu,\thinspace\nu,\ldots)$ in the
above relations except for Eq.\ (\ref{eq:coh-92}) where we need the
Lorentz indices to explicitly express the relation.

By writing the amplitude (\ref{eq:coh-79}) in terms of $(C_{a}^{(q)},\thinspace\overline{D}_{a}^{(q)})$
and the form factors, and comparing it with Eq.\ (\ref{eq:coh-57}),
we obtain
\begin{equation}
\left(C_{S},\thinspace\overline{D}_{S}\right)=\sum_{q=u,d}n_{q}\left(C_{S}^{(q)}f_{SS}+\overline{D}_{S}^{(q)}f_{PP},\thinspace C_{S}^{(q)}f_{SP}+\overline{D}_{S}^{(q)}f_{PS}\right),\label{eq:coh-93}
\end{equation}
\begin{equation}
\left(C_{P},\thinspace\overline{D}_{P}\right)=\sum_{q=u,d}n_{q}\left(C_{P}^{(q)}f_{PS}-\overline{D}_{S}^{(q)}f_{SP},\thinspace-C_{P}^{(q)}f_{PP}+\overline{D}_{P}^{(q)}f_{SS}\right),\label{eq:coh-93-1}
\end{equation}
\begin{equation}
\left(C_{V},\thinspace i\overline{D}_{V}\right)=\sum_{q=u,d}n_{q}\left(C_{V}^{(q)}f_{VV}+i\overline{D}_{V}^{(q)}f_{AA},\thinspace C_{V}^{(q)}f_{VA}+i\overline{D}_{V}^{(q)}f_{AV}\right),\label{eq:coh-93-2}
\end{equation}
\begin{equation}
\left(C_{A},\thinspace i\overline{D}_{A}\right)=\sum_{q=u,d}n_{q}\left(C_{A}^{(q)}f_{AV}+i\overline{D}_{A}^{(q)}f_{VA},\thinspace C_{A}^{(q)}f_{AA}+i\overline{D}_{A}^{(q)}f_{VV}\right),\label{eq:coh-93-2-1}
\end{equation}
\begin{equation}
\left(C_{T},\thinspace\overline{D}_{T}\right)=\sum_{q=u,d}n_{q}\left(C_{T}^{(q)}f_{T}-\overline{D}_{T}^{(q)}f_{T'},\thinspace C_{T}^{(q)}f_{T'}+\overline{D}_{T}^{(q)}f_{T}\right).\label{eq:coh-93-3}
\end{equation}
These are the relations that connect the nucleus
couplings $(C_{a},\thinspace\overline{D}_{a})$
and the quark couplings $(C_{a}^{(q)},\thinspace\overline{D}_{a}^{(q)})$.

\bibliographystyle{apsrev4-1}
\bibliography{ref}

\end{document}